%% file: main.tex
\documentclass[twocolumn]{aastex631}

\usepackage{acro}
\usepackage{hyperref}
\usepackage{xspace}

\acsetup{single=true, list/sort=true, cite/cmd=\citep, cite/group=true, cite/group/cmd=\citealt}
\acsetup{patch/longtable=false}

\newcommand\rearth{R$_{\rm \oplus}$\xspace}
\newcommand\mjup{M$_{\rm Jup}$\xspace}
\newcommand\vespa{\texttt{vespa}\xspace}
\newcommand\triceratops{\texttt{TRICERATOPS}\xspace}

\newcommand\vp{23\xspace}
\newcommand\falsep{29\xspace}
\newcommand\lp{15\xspace}
\newcommand\pfp{18\xspace}
\newcommand\confirmed{14\xspace}
\newcommand\inconclusive{4\xspace}
\newcommand\marginal{33\xspace}

\DeclareAcronym{TSM}{short=TSM, long=Transmission Spectroscopy Metric}
\DeclareAcronym{ESM}{short=ESM, long=Emission Spectroscopy Metric}
\DeclareAcronym{TESS}{short=TESS, long=the Transiting Exoplanet Survey Satellite}
\DeclareAcronym{JWST}{short=JWST, long=JWST}
\DeclareAcronym{TOI}{short=TOI, long=TESS Object of Interest, long-plural-form=TESS Objects of Interest}
\DeclareAcronym{TOIs}{short=TOIs, long=TESS Objects of Interest}
\DeclareAcronym{TFOP}{short=TFOP, long=the TESS Follow-up Observation Program}
\DeclareAcronym{TIC}{short=TIC, long=the TESS Input Catalog}
\DeclareAcronym{SPOC}{short=SPOC, long=Science Processing Operations Center}
\DeclareAcronym{QLP}{short=QLP, long=Quick Look Pipeline, cite={huang2020qlplightcurves}}
\DeclareAcronym{DAVE}{short=\texttt{DAVE}, long=the Discovery And Vetting of Exoplanets, cite={kostov2019adave}}
\DeclareAcronym{ZTF}{short=ZTF, long=the Zwicky Transient Facility, cite={bellm2019ztf}}
\DeclareAcronym{PDCSAP}{short=PDC-SAP, long=Presearch Data Conditioning Simple Aperture Photometry, cite={smith2012kepler, stumpe2012kepler, stumpe2014multiscale}}
\DeclareAcronym{FPP}{short=FPP, long=false positive probability}
\DeclareAcronym{NFPP}{short=NFPP, long=nearby false positive probability}

\begin{document}

\title{Identification of the Top TESS Objects of Interest for Atmospheric 
 Characterization of Transiting Exoplanets with JWST}

\input{authors}

\correspondingauthor{Benjamin J. Hord}
\email{benhord@astro.umd.edu}

\begin{abstract}

JWST has ushered in an era of unprecedented ability to characterize exoplanetary atmospheres. While there are over 5,000 confirmed planets, more than 4,000 TESS planet candidates are still unconfirmed and many of the best planets for atmospheric characterization may remain to be identified. We present a sample of TESS planets and planet candidates that we identify as ``best-in-class'' for transmission and emission spectroscopy with JWST. These targets are sorted into bins across equilibrium temperature $T_{\mathrm{eq}}$ and planetary radius $R{_\mathrm{p}}$ and are ranked by transmission and emission spectroscopy metric (TSM and ESM, respectively) within each bin. In forming our target sample, we perform cuts for expected signal size and stellar brightness, to remove sub-optimal targets for JWST. Of the 194 targets in the resulting sample, 103 are unconfirmed TESS planet candidates, also known as TESS Objects of Interest (TOIs). We perform vetting and statistical validation analyses on these 103 targets to determine which are likely planets and which are likely false positives, incorporating ground-based follow-up from the TESS Follow-up Observation Program (TFOP) to aid the vetting and validation process. We statistically validate \vp TOIs, marginally validate \marginal TOIs to varying levels of confidence, deem \falsep TOIs likely false positives, and leave the dispositions for \inconclusive TOIs as inconclusive. \confirmed of the 103 TOIs were confirmed independently over the course of our analysis. We provide our final best-in-class sample as a community resource for future JWST proposals and observations. We intend for this work to motivate formal confirmation and mass measurements of each validated planet and encourage more detailed analysis of individual targets by the community.

\end{abstract}

\section{Introduction} \label{sec:intro}

Since the first exoplanets were discovered by \cite{wolszczan1992firstexoplanet} and \cite{mayor1995jupiter}, over 5,000 exoplanets have been confirmed, opening up a wide array of planets of varying sizes, temperatures, and masses for study. The rate of exoplanet discovery has notably accelerated over time, originating with serendipitous or targeted observations and culminating in the concerted efforts of ground-based surveys such as the Wide Angle Search for Planets \citep[WASP;][]{pollacco2006wasp}, the Hungarian-made Automated Telescope Network \citep[HATNet;][]{bakos2004wide}, and HATSouth \citep{Bakos:2013} and space-based observatories such as  the COnvection, ROtation and planetary Transits satellite \citep[CoRoT;][]{auvergne2009corot,moutou2013corot}, Kepler \citep{borucki2010keplerintro}, K2 \citep{howell2014k2intro}, and the Transiting Exoplanet Survey Satellite (TESS, \citealt{ricker2015tessintro}).

Although the exoplanet discovery process can reveal important properties of planets like mass and radius, further observations and analysis are required to understand the conditions on the planets themselves and examine the planet's atmospheric composition and dynamics. The first observation of an exoplanetary atmosphere was conducted by \citet{charbonneau2002firstatmosphere}, and since then, in a parallel to the diversity of the types of exoplanets, spectroscopic characterization has revealed a wide variety of atmospheric compositions and aerosol properties as well \citep[e.g.,][]{sing2016hjclouds,welbanks2019massmetallicity,mansfield2021hjcomposition,changeat2022hjeclipses,august2023wasp-77bmetallicity}. 

Transmission and emission spectroscopy have proven to be the workhorses of exoplanetary atmospheric characterization. These methods utilize the absorption of stellar flux transmitted through the exoplanetary atmosphere and the thermal emission from the exoplanet to probe the atmospheric characteristics of the planet. Exoplanet atmospheric characterization and spectral modeling have greatly expanded our understanding of the formation and evolution of planets, the physical and chemical processes that shape planetary atmospheres, and atmospheric aerosol properties \citep[e.g.][]{madhusudhan2019atmospherechallenges, molliere2022formationandatmospheres, wordsworth2022rockyatmos} as well as the range of diverse conditions within each of these individual topics. As the outermost layer of a planet, the atmosphere is the easiest component of an exoplanet to probe in detail and can be used to infer other planetary properties.

Although space- and ground-based resources for atmospheric characterization have become more abundant since the first transmission spectrum was taken, these resources remain in high demand. The premier atmospheric characterization tools have largely been the Hubble Space Telescope and, until its retirement in 2020, the Spitzer Space Telescope, both of which have historically been heavily oversubscribed. High-resolution spectrographs on ground-based telescopes have become increasingly important in the study of exoplanet atmospheres, but these are often limited by what is visible in the night sky and signal-to-noise ratios.

The highly-anticipated \acs{JWST} launched in 2021 \citep{gardner2006jwstintro, gardner2023jwst} with promises of greatly improved capabilities for transit and eclipse exoplanet atmospheric characterization \citep[e.g.][]{deming2009jwstfollowup,greene2016jwstexoatmospheres,stevenson2016jwsters} owing to its large aperture and infrared (IR) instrument complement. Although still early in its mission, \ac{JWST} has already begun delivering on these promises with its first year of exoplanet results \citep[e.g.][]{tsai2023wasp-39bso2, ahrer2023wasp39jwsters,greene2023trappist1bemissionspec,kempton2023gj1214bphasecurve}. This is not even to mention \ac{JWST}'s capabilities for the spectroscopy of directly imaged exoplanets \citep[e.g.,][]{miles2023jwstdirectimageers} which is impressive but outside the scope of this work. But time on \ac{JWST} is in high demand, and this, coupled with the review process for general observer programs, has resulted so far in a patchwork of exoplanet atmospheric observations.

\begin{figure*}
    \centering
    \includegraphics[width=0.8\textwidth]{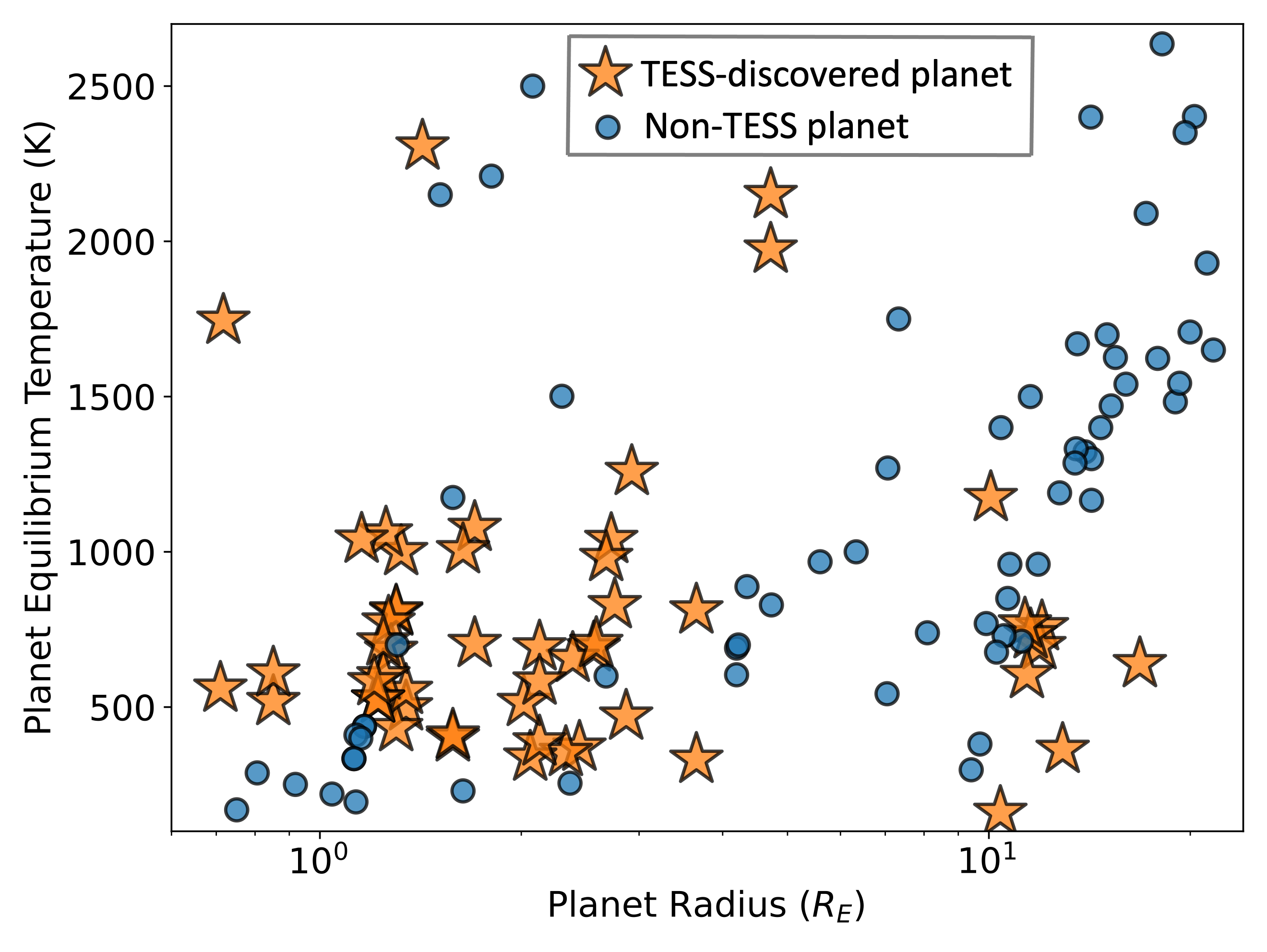}
    \caption{All of the \ac{JWST} exoplanet targets that are approved for transmission or emission spectroscopy observations in Cycles 1 and 2 across planetary equilibrium temperature and radius. Yellow stars represent approved \ac{JWST} targets that were discovered by \acs{TESS} while those represented by blue circles represent planets not discovered by \acs{TESS}. As evidenced by the plot, \acs{TESS}-discovered planets constitute a large proportion of approved \ac{JWST} Cycle 1 and 2 targets for transmission or emission spectroscopy and cover a wide range of parameter space.}
    \label{fig:jwst_tess_targets}
\end{figure*}

When it comes to identifying targets for atmospheric characterization observations, there is a critical synergy between JWST and \acs{TESS}. Touted from the very beginning as a ``finder scope for \ac{JWST}," the almost-all sky survey strategy of \ac{TESS} was intended to find a myriad of new planets around bright, nearby stars that would be amenable to atmospheric characterization with \ac{JWST} \citep{deming2009jwstfollowup}, in contrast to the dimmer, more distant host stars of Kepler planetary systems. So far, \ac{TESS} has discovered more than 300 confirmed planets, with more than 4,000 planet candidates classified as unconfirmed \acp{TOI} without either a false positive or confirmed planet disposition$\footnote{\url{https://exoplanetarchive.ipac.caltech.edu/docs/counts\_detail.html}}$. There is currently no published false positive rate for \ac{TESS}, although recent work estimates that it could be somewhere between 15$\%$ and 47$\%$ depending on the mass of the planet and host star \citep{zhou2019tesshjoccurrencerates,kunimoto2022tessyield}. Therefore, it is probably that many of these 4,000 \acp{TOI} are false positives. However, if even a fraction of them are true planets, this would dramatically grow the sample of planets whose atmospheres may be well-suited to observe and characterize with \ac{JWST}.

In fact, some of the highest quality (i.e.\ highest signal-to-noise) atmospheric characterization exoplanet targets likely still lie among the unconfirmed \acp{TOI} list, since \ac{TESS} has unique capabilities for finding small planets orbiting bright stars in particular. The \ac{JWST}-\ac{TESS} synergy is demonstrated especially by the fact that $\sim$37$\%$ of \ac{JWST} Cycle 1 and $\sim$56$\%$ of \ac{JWST} Cycle 2 exoplanet targets are \ac{TESS} discoveries. This high proportion of \ac{TESS}-discovered \ac{JWST} targets is displayed in Figure \ref{fig:jwst_tess_targets}. With \ac{JWST} already flying, it is of the utmost importance to systematically and expeditiously identify the best \ac{JWST} targets to provide a uniform coverage of parameter space.

In an effort to better streamline use of \ac{JWST} for atmospheric characterization and identify which targets are likely to exhibit the most clearly detectable features in their atmospheric spectra, we present a set of ``best-in-class'' targets for transmission and emission spectroscopy. Our best-in-class sample consists of the targets ranked in the top five according to the \ac{TSM} and \ac{ESM} from \cite{kempton2018tsmesm} within each cell of a grid spanning the $R_{\mathrm{p}}$-$T_{\mathrm{eq}}$ space, which is described in Section \ref{sec:grid}. $R_{\mathrm{p}}$-$T_{\mathrm{eq}}$ axes were chosen since radius is expected to be a proxy for metallicity \citep{loggers1998mass-magnitude,fortney2013lowmassplanetatmos} while temperature correlates to chemistry and aerosol formation \citep{gao2020aerosols} and both parameters are easy to estimate for transiting exoplanets. Metallicity and atmospheric chemistry can both provide insights to the formation, physical processes, and composition of a planet's atmosphere and are important to probe. We account for the technical capabilities of \ac{JWST}'s instruments through the inclusion and calculation of various additional metrics (e.g.\ stellar host magnitude, expected atmospheric signal size, and observability metrics benchmarked against JWST's instrumental capabilities) for each target and further incorporate these values into our rankings, thus tuning our best-in-class sample to \ac{JWST} specifically.

In our rankings, we initially make no distinction between confirmed planets and unconfirmed \acp{TOI} in order to assess how the TESS planet candidates fit in with the overall sample and to identify which \acp{TOI} might displace known planets as best-in-class atmospheric characterization targets. For each unconfirmed \ac{TOI} on our best-in-class list, we perform cursory vetting and statistical validation to determine which targets are likely false positives and which are worthy of additional follow-up prior to future atmospheric characterization observations with \ac{JWST}. We note that while we only statistically ``validate'' planets rather than label them as ``confirmed'', we consider them to be planets for the purposes of our best-in-class sample \citep{torres2004testingfalsepositives,torres2011falsepositives}. Our aim is to produce a sample of planets (or likely planets) well-suited for \ac{JWST} atmospheric characterization to serve as a community resource for upcoming \ac{JWST} proposal cycles and future observing programs aimed at regions of planetary parameter space where the highest SNR targets have yet to be identified. Under mass assumptions that we describe in Section \ref{sec:grid}, these targets are expected to be well-suited for \ac{JWST}.

In Section \ref{sec:grid}, we outline our methodology for obtaining our best-in-class sample including the data origin, the metrics calculated, and the specific boundaries in parameter space that were used when defining each class of planets. In Section \ref{sec:observations}, we describe the follow-up observations obtained to aid in our vetting and validation analyses of each unconfirmed \ac{TOI} contained in our best-in-class sample. Section \ref{sec:vetting} details our vetting procedures, the follow-up and independent resources that were used in our consideration of false positive scenarios for each unconfirmed \ac{TOI}, and the criteria against which each target was compared. Section \ref{sec:validation} walks through our statistical validation procedures including our implementation of statistical validation software and the disposition categories that we sorted each unconfirmed \ac{TOI} into based on the results of our vetting and validation analyses. In Section \ref{sec:results} we summarize the results of our vetting and statistical validation including which unconfirmed \acp{TOI} were statistically validated and which we considered likely false positives. Our findings are summarized in Section \ref{sec:summary}.

\section{Grid Generation} \label{sec:grid}

\begin{figure*}[!h]
    \centering
    \includegraphics[width=0.95\textwidth]{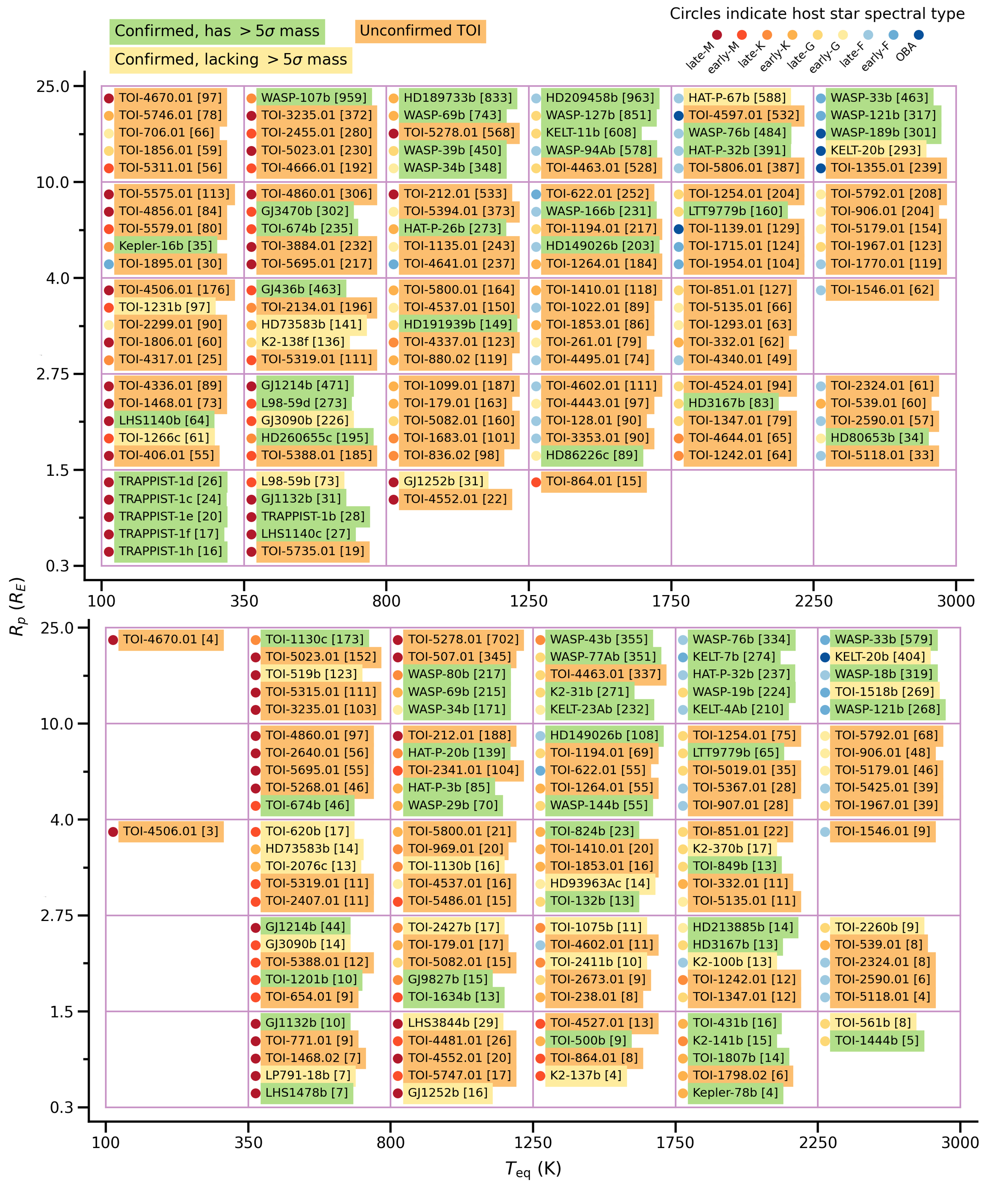}
    \caption{Our best-in-class targets for transmission (\textit{top}) and emission (\textit{bottom}) spectroscopy as of November 3, 2022 sorted by equilibrium temperature, $T_{\rm eq}$, and planetary radius, $R_{\rm p}$. Target names are shown with the respective spectroscopy metrics (ESM or TSM) in brackets next to the name. Targets are sorted within each cell by spectroscopy metric in descending order. Approximate stellar type of the host star is denoted by the colored circle to the left of each name, determined by the reported effective temperature. Targets are color-coded by mass status: green targets are confirmed planets with mass measurements $>$5$\sigma$, yellow targets are confirmed planets with mass measurements $<$5$\sigma$, and orange targets are unconfirmed \acp{TOI}.}
    \label{fig:original_grids}
\end{figure*}

\begin{figure*}
    \centering
    \includegraphics[width=0.8\textwidth]{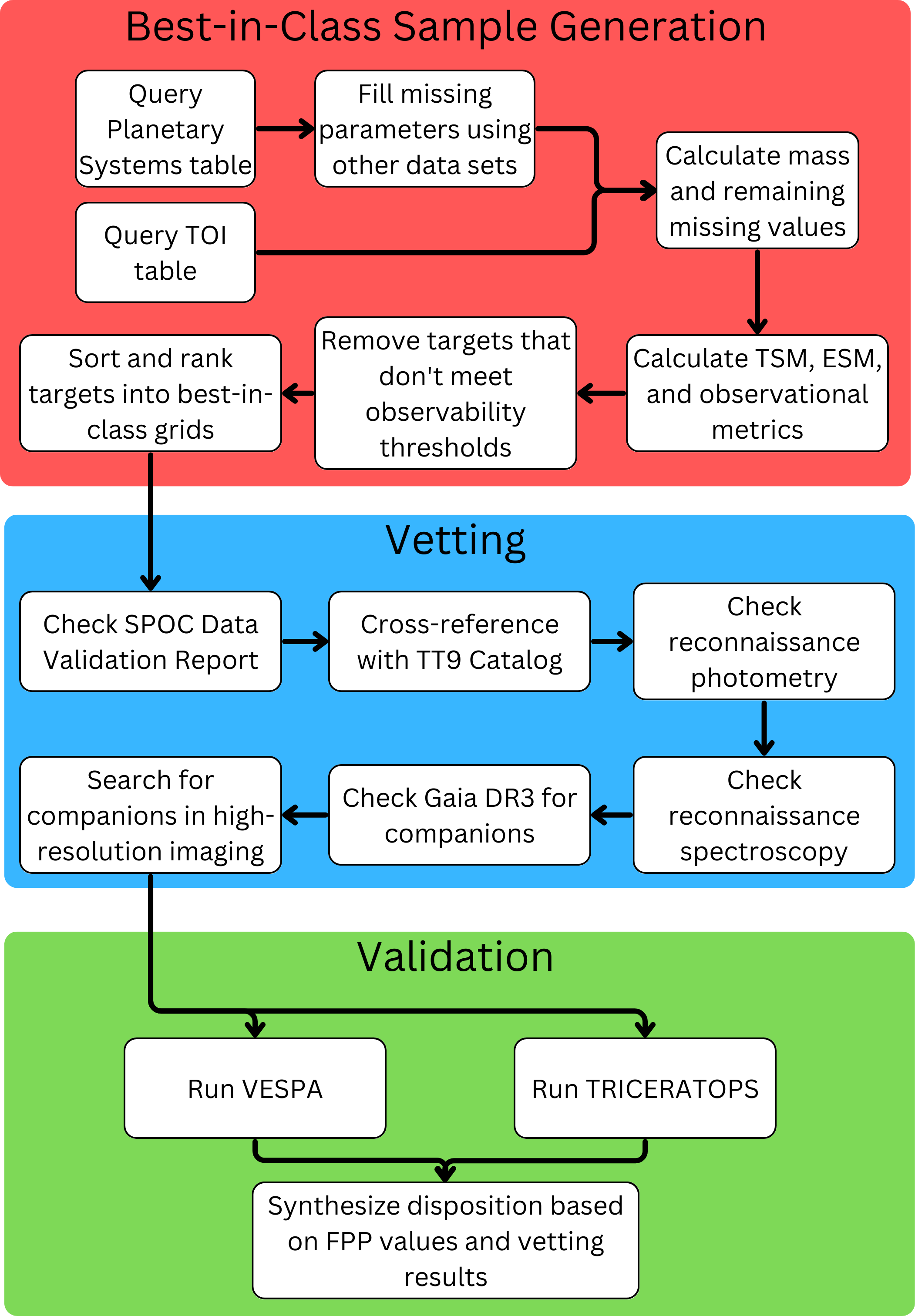}
    \caption{A schematic outline of our analysis procedure. From the initial query of the Exoplanet Archive and generation of the best-in-class sample, each target went through every step of the procedure to check for factors that could indicate a false positive to arrive at a final disposition. Not every vetting step applied to every target due to lack of follow-up, so each vetting step was applied when possible but skipped when not.}
    \label{fig:vetting_procedure}
\end{figure*}

Identifying targets across $R_{\mathrm{p}}$-$T_{\mathrm{eq}}$ space that are well-suited to atmospheric characterization with \ac{JWST} is critical to our understanding of exoplanet atmospheres. By sampling across this parameter space, we expect to cover a range of metallicities as well as atmospheric chemistry and aerosol regimes that would allow us to tease out trends and test models on the population level. This could include a mass-metallicity relation, an aerosol-$T_{\mathrm{eq}}$ relation, or a transition between planets that have CO vs. CH$_{4}$ in their atmospheres as the dominant carbon carrier. To accomplish this, we've divided up the $R_{\mathrm{p}}$-$T_{\mathrm{eq}}$ parameter space into a grid, sorted each planet and planet candidate into cells within this grid, and ranked each target according to its expected signal-to-noise ratio approximated via its \ac{TSM} or \ac{ESM}. The samples for both transmission and emission spectroscopy can be found in Figure \ref{fig:original_grids} and a visual outline of our sample generation is shown in the top box of Figure \ref{fig:vetting_procedure}.

\subsection{Provenance of Sample Parameters and TSM $\&$ ESM Calculation} \label{ssec:sample}

In order to obtain a standardized list of planets and planet candidates to consider when determining which are the best-in-class for atmospheric characterization with \ac{JWST}, we relied on the data tables maintained by the NASA Exoplanet Archive\footnote{\url{https://exoplanetarchive.ipac.caltech.edu/}} and the parameter values contained therein. The Exoplanet Archive collates parameter sets for confirmed and unconfirmed planets and acts as a single repository for published parameter values for each target. For the confirmed planets, we downloaded the Planetary Systems table which contains every planet that has a published validation or confirmation and the accompanying set of parameter values with a single parameter set labeled as the default by the archive staff for each planet. For the unconfirmed \acp{TOI}, we downloaded the \ac{TESS} Candidates table from the Exoplanet Archive, which updates directly from the \ac{TESS} \ac{TOI} Catalog \citep{guerrero2021tess} with new targets and refined parameter values from the \ac{TESS} mission. These two tables were both downloaded on November 3, 2022. The highest \ac{TOI} number alerted at this time was TOI-5863.

We elected to use the parameter set denoted as the default set of values for each of the planets in the Planetary Systems table throughout our analysis. In the case that the default parameter set was incomplete and missing values for critical parameters necessary to our analysis, values were pulled from other, non-default parameter sets for each planet, if they existed. Critical values included $R_{\rm p}$, $R_{*}$, $T_{*}$, $a$, $J$ magnitude, and $K$ magnitude. Values with lower uncertainties from other parameter sets were given priority for inclusion in the final parameter set. 

\begin{figure*}
    \centering
    \includegraphics[width=0.99\textwidth]{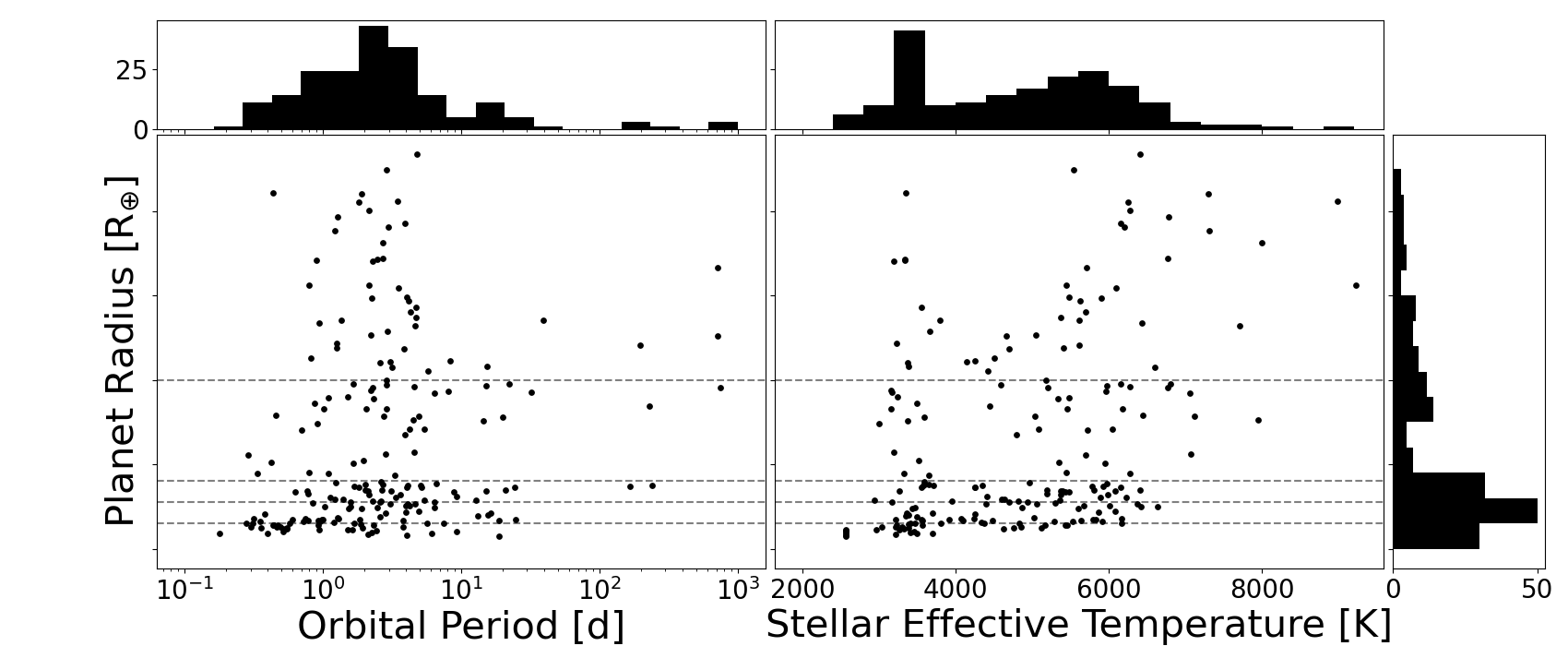}
    \caption{The spread of targets in our best-in-class samples. \textit{Left:} the orbital periods and planetary radii of the combined \ac{TSM} and \ac{ESM} best-in-class samples. \textit{Right:} the effective temperature of the host stars and the planetary radii of the same combined best-in-class sample. Also shown are the marginal distributions for each variable. The edges of the radius bins are represented by the gray dashed horizontal lines. Although only selected to adequately cover the planetary radius and equilibrium temperature parameter space, the best-in-class sample exhibits good coverage of multiple different parameter spaces and can be considered a representative subset of exoplanetary targets.}
    \label{fig:sample_scatter}
\end{figure*}

We calculated the \ac{TSM} and \ac{ESM} for each planet according to the prescription outlined in \cite{kempton2018tsmesm}, specifically equations 1 and 4. The calculation of \ac{TSM} and \ac{ESM} assumes cloud-free atmospheres, solar composition for planets larger than 1.5 \rearth, and a pure H$_{2}$O steam atmosphere for planets smaller than 1.5 \rearth. These two values represent analytical metrics that quantify the expected signal-to-noise in transmission and thermal emission spectroscopy for a given planet and can be used to identify which planets are best-suited for atmospheric characterization with \ac{JWST} relative to one another. We maintained two separate samples for best-in-class targets: one for transmission spectroscopy driven by \ac{TSM} and the other for emission spectroscopy driven by \ac{ESM}. Both of these initially started with the same overall sample of planets and planet candidates downloaded from the Exoplanet Archive and were each shaped by the observational constraints unique to each respective sample. Figure \ref{fig:sample_scatter} illustrates the parameter space coverage of our combined best-in-class samples.

Even after pulling values from other parameter sets, some targets did not contain finite values for all of the parameters necessary to calculate the spectroscopy metrics and the observability criteria with which we defined and ranked our sample. For targets without a value for the ratio between semi-major axis and stellar radius, $a$/$R_{*}$, we converted both the semi-major axis $a$ and the stellar radius $R_{*}$ to units of meters and took the ratio of the two. In the case that $a$ was missing but $a$/$R_{*}$ was a finite value, $a$/$R_{*}$ was multiplied by $R_{*}$ to calculate $a$. A similar procedure was performed for the ratio between the planet and stellar radii, $R_{\mathrm{p}}$/$R_{\mathrm{*}}$. We preferred to use the reported ratios if they existed to reduce the propogation of potential errors in generating these ratios from the reported values of their individual components. Reported mass and equilibrium temperature values were used when reported, but were calculated later in the procedure if unavailable. All targets that still lacked full parameter sets to perform the necessary calculations were removed from the sample. We checked each parameter set to ensure that $R_{\rm p}$/R$_{*}$ $<$ 1 and targets with values that did not conform to this criterion were replaced with a value from another parameter set, if available.

For planets from the Planetary Systems table and candidates from the TOI list that did not have published masses, we calculated masses using a mass-radius distribution adapted from the mean of the \cite{chen2018massradius} mass-radius distribution. Specifically, we set the S$^{3}$ coefficient to be 0.01 rather than -0.044, to ensure that each radius value corresponded to a unique mass, while minimally affecting the shape of the curve as presented in \cite{chen2018massradius}. We used this distribution up to planetary radii of 15~\rearth, fixing the mass of planets larger than this threshold to 1 \mjup. Above this radius, the scatter of the mass-radius distribution is large and results in a mean that is nearly constant in mass across radius. This is the same procedure that is used by the Exoplanet Archive to calculate expected masses. 

We divided the sample into three categories: confirmed planets with $>$ 5$\sigma$ mass measurements, planets marked as confirmed on the Exoplanet Archive with $<$ 5$\sigma$ mass measurements, and unconfirmed planet candidates without any mass measurement. \citet{batalha2019massprecision} showed that different mass confidence levels result in different precision with which an exoplanet's atmosphere can be characterized. A stratification of these targets based on mass measurement will also allow the community to better prioritize follow-up resources for the best-in-class targets and allowed us to identify which targets are unconfirmed and in need of statistical validation.

Additionally, we calculated the mass of the host star for each \ac{TOI} based on the star's log $g$ and radius because stellar mass is not included in the Exoplanet Archive's \ac{TOI} table. Using the host star's reported effective temperature, we also assigned each host star an approximate stellar type for reference. We then calculated the equilibrium temperatures $T_{\mathrm{eq}}$ of each planet -- both \ac{TOI} and confirmed -- according to Equation 3 of \cite{kempton2018tsmesm}. This was done to ensure a uniform data set for $T_{\rm eq}$ since the definition of equilibrium temperature varies with each data set on the Exoplanet Archive, with different assumptions regarding surface albedo and atmospheric heat distribution serving as variables with no set standard. Since $T_{\rm eq}$ is integral to our determination of the best targets for transmission and emission spectroscopy, we elected to calculate the value for each planet and planet candidate to ensure a uniform comparison. Our calculation of $T_{\rm eq}$ assumes zero albedo and full day-night heat redistribution.

\subsection{Observability Cuts} \label{ssec:observability_cuts}

While useful for relative comparisons between targets, the \ac{TSM} and \ac{ESM} only predict the signal to noise but do not account for other observability considerations such as the absolute signal size relative to the instrumental noise floor or the target being within an instrument's brightness limits. To incorporate the observability of our sample with \ac{JWST} into our best-in-class rankings, we also calculated the expected sizes of transmission spectral features and secondary eclipse depth for transmission and emission spectroscopy, respectively. 

\subsubsection{Observability in Transmission} \label{sssec:transmiss_observability}

We again follow the prescription outlined in \cite{kempton2018tsmesm}, expressing the size of expected spectral features at one scale height as 

\begin{equation}
    \frac{2R_{\rm p}}{R_{*}^{2}} \times \frac{k T_{\rm eq}}{\mu g}
    \label{eq:tran_signal}
\end{equation}

\noindent where $R_{\mathrm{p}}$ is the planetary radius, $R_{\mathrm{*}}$ is the radius of the host star, $k$ is the Boltzmann constant, $T_{\mathrm{eq}}$ is the equilibrium temperature of the planet, $\mu$ is the mean molecular weight of the atmosphere, and $g$ is the surface gravity of the planet. For planets with $R_{\mathrm{p}}$ $>$ 1.5 \rearth, we assume $\mu$ = 2.3 (in units of proton mass, m$_{p}$) while for planets with $R_{\mathrm{p}}$ $<$ 1.5 \rearth, we assume $\mu$ = 18 proton masses, following the assumption that all planets in a given radius bin have the same atmospheric composition as made by \cite{louie2018simjwstspec}. We calculated $g$ using the expression $g$=$GM_{\mathrm{p}}$/$R_{\mathrm{p}}^2$ where $G$ is the gravitational constant and $M_{\mathrm{p}}$ and $R_{\mathrm{p}}$ are the mass and radius of the planet, respectively. The second term of Equation \ref{eq:tran_signal} represents the scale height of the planetary atmosphere, $H$. This is used as a proxy for spectral feature size as it represents the depth into that atmosphere that is probed at a specific wavelength, which in turn determines the measured wavelength-dependent differences in transit depth.

We assumed a depth of 2$H$ when calculating expected spectral feature size based off of the spread in the sizes of H$_{2}$O features observed using the Hubble Space Telescope's NIR WFC3 instrument \citep{stevenson2016clouds}. The average size of these features was reported to be $\sim$1.5$H$, but at longer wavelengths such as those probed by \ac{JWST}, the size of spectral features for molecules such as H$_{2}$O increases \citep[e.g.,][]{coulombe2023broadband}, so we elected to assume a depth slightly above the average reported by \cite{stevenson2016clouds}. Assuming a larger expected spectral feature also allows for us to capture more planets for comparison within our sample as well as to account for differences in cloud cover or the mean molecular weight of exoplanet atmospheres.

In fact, for all constraints applied to our sample, we chose liberal thresholds in order to allow for more targets to appear in our best-in-class sample, especially in parameter spaces where there otherwise would be no promising targets. This was done not only for illustrative purposes, but also to attempt to account for some of the variance in parameters governing exoplanet atmospheres and potentially improved observational capabilities going forward.

To ensure that all best-in-class targets would be observable with \ac{JWST}, we imposed a requirement for a 2$\sigma$ spectral signal size assuming a noise floor of 10 ppm for the NIRCam, NIRISS, and NIRSpec instruments on \ac{JWST}. These instruments are all ideal for transmission spectroscopy since their wavelength coverage includes prominent transmission spectral features. We note the \ac{TSM} was benchmarked for use with NIRISS \citep{kempton2018tsmesm}.

\subsubsection{Observability in Emission} \label{sssec:emiss_observability}

We perform a similar procedure for the secondary eclipse depth in order to determine which targets are amenable for emission spectroscopy with the MIRI instrument onboard \ac{JWST}. The expected secondary eclipse depth can be estimated using the expression

\begin{equation}
    \frac{B_{7.5}(T_{\rm day})}{B_{7.5}(T_{*})} \times \left( \frac{R_{\rm p}}{R_{*}} \right)^{2}
    \label{eq:ec_depth}
\end{equation}

\noindent where $B_{\mathrm{7.5}}$ is the Planck function evaluated for a given temperature at a representative wavelength of 7.5 $\mu$m, $T_{\mathrm{day}}$ is the dayside temperature of the planet as calculated by 1.1 $\times$ $T_{\mathrm{eq}}$, $T_{\mathrm{*}}$ is the effective temperature of the host star, and $R_{\mathrm{p}}$/$R_{\mathrm{*}}$ is the ratio of the planetary and stellar radii. We calculate the dayside temperature as 1.1 $\times$ $T_{\mathrm{eq}}$ to account for the dayside hotspot on the planet, following the analysis by \cite{kempton2018tsmesm} that tuned this relation according to a suite of global circulation and 1D atmospheric models. The 7.5 $\mu$m was chosen as the representative wavelength since it is the center of the ``conservative'' MIRI LRS bandpass on \ac{JWST} as data beyond 10 $\mu$m are often unreliable \citep{bell2023wasp43bmiri,kempton2023gj1214bphasecurve} and 7.5 $\mu$m is still near the peak of the MIRI LRS response function \citep{rieke2015miri, kendrew2015miri}. We imposed a requirement that the secondary eclipse depth be measurable to the 3$\sigma$ level assuming a noise floor of 20 ppm for the MIRI instrument on \ac{JWST}. There were more small planets contained within the emission spectroscopy sample and so we were able to adopt a more conservative 3$\sigma$ threshold rather than the 2$\sigma$ threshold applied to the transmission spectroscopy sample. We also imposed an \ac{ESM} $>$ 3 requirement on our emission spectroscopy sample to remove targets that would produce small secondary eclipses even under ideal observing conditions with \ac{JWST}. Like \ac{TSM} with NIRISS, \ac{ESM} was benchmarked for use with MIRI, which is ideal for emission spectroscopy among \ac{JWST}'s instruments thanks to its longer wavelength coverage that maximizes the ratio between the flux of the planet and that of the host star.

\subsection{Additional Cuts and Organizing the Sample} \label{ssec:add_cuts_org_sample}

We applied additional observability cuts to the sample to ensure that each of our best-in-class targets would be observable by \ac{JWST} and would produce significant spectral detections. For transmission spectroscopy targets, we restricted the $J$ magnitude of the host star to $>$ 6.0 while for emission spectroscopy targets we restricted the $K$ magnitude of the host stars to $>$ 6.4. These values represent the approximate maximum brightnesses at which the NIRCam long-wavelength channel grism spectroscopy \citep[which can observe the brightest stars of the near-infrared spectroscopic modes,][]{beichman14} and MIRI Low Resolution Spectroscopy \citep[LRS,][]{kendrew16} modes will not saturate, respectively, according to v2.0 of the JWST exposure time calculator \citep{pontoppidan16}. We also removed any planets or planet candidates with impact parameter $b$ $>$ 0.9 to remove grazing transits that could produce unreliable transit depths.

We then divided our full sample of targets that are observable with \ac{JWST} into bins of planetary radius and equilibrium temperature to determine which targets are best for atmospheric characterization in their class. This division included both confirmed planets and unconfirmed planet candidates. The edges of these bins in planetary radius were chosen in order to match the cutoffs used in \cite{kempton2018tsmesm}, setting the minimum and maximum radii to include the smallest and largest transiting planets at the time the Exoplanet Archive was queried. The temperature bin edges were chosen to capture the ultra-hot Jupiters at $T_{\rm eq}$ $>$ 2250 K, the carbon equilibrium chemistry transition from CO (and CO$_{2}$) to CH$_{4}$ around 800 K \citep[assuming an otherwise solar C/O ratio,][]{fortney2013lowmassplanetatmos}, and roughly equal spacing otherwise. The coldest temperature bin in our sample was chosen to encompass the habitable zone.

\subsection{Description of Best-in-Class Grids} \label{ssec:grid_description}

The planets contained within each bin in radius and temperature space were then sorted and ranked by \ac{TSM} and \ac{ESM} for the transmission spectroscopy and emission spectroscopy samples, respectively. This ranking was agnostic to confirmation status and the existence of a well-constrained mass, resulting in a combination of confirmed planets and unconfirmed planet candidates within each grid cell. The top five targets in each bin are considered the best-in-class for that portion of parameter space. Our rankings of the transmission and emission spectroscopy targets are contained within the grids shown in Figure \ref{fig:original_grids}.

Almost every bin for both the transmission and emission target samples has at least one unconfirmed planet candidate, with most bins dominated by unconfirmed candidates. While certainly not all of the planet candidates are true planets, if even a fraction of the them are, these rankings indicate that there is a large number of \ac{TESS} planet candidates that are both (i) among the best currently known targets for atmospheric characterization with \ac{JWST} from a signal-to-noise ratio perspective; and (ii) required to provide a uniform coverage of the $R_{\mathrm{p}}$-$T_{\mathrm{eq}}$ space.

\section{Follow-up Observations} \label{sec:observations}

In order to determine which of the \ac{TESS}-discovered planet candidates in our best-in-class samples are true planets, we first collated all of the follow-up observations for each target. These follow-up observations provided valuable, independent information on the validity of each planet candidate as true planets. We worked closely with \ac{TFOP}\footnote{\url{https://tess.mit.edu/followup}} subgroups (SGs) to compile available photometric, spectroscopic, and imaging follow-up observations for each target. These observations were used in initial vetting to determine whether each target was a likely false positive or if it could proceed to more in-depth vetting and validation. \ac{TFOP} follow-up observations and the constraints that they impose on the system were incorporated into our vetting and statistical validation procedures where possible (see Sections \ref{sec:vetting} and \ref{sec:validation}). The follow-up resources used in vetting and validating the best-in-class planet candidates are summarized here with a representative sample of the specific observations used for individual targets detailed in Table \ref{tab:tfop_obs} located in Appendix \ref{app:tfop_obs} and a full, machine-readable version available from the online version of this article. An outline of where follow-up observations were used in our vetting procedures can be found in the middle panel of Figure~\ref{fig:vetting_procedure}.

\subsection{Ground-based Photometry} \label{ssec:sg1_photometry}

\ac{TFOP}'s Sub Group 1 \citep[SG1;][]{collins:2019} performed ground-based photometry for almost all of the targets in our best-in-class samples in order to clear the background fields of eclipsing binaries (EBs), to check if the candidate transit signal could be identified as on target, and to check the chromaticity of the transit shape and depth. This ground-based photometry was taken by a variety of observatories over a span of multiple years. The {\tt TESS Transit Finder}, which is a customized version of the {\tt Tapir} software package \citep{Jensen:2013}, was used to schedule the transit follow-up observations included here. Below we detail the observatories, instruments, and data reduction methods used to obtain the ground-based photometry for our samples. Unless otherwise noted, all image data were calibrated and photometric data were extracted using {\tt AstroImageJ} \citep{Collins:2017}. Further discussion on the use of ground-based photometry in vetting and validation can be found in Sections \ref{ssec:phot_vet} and \ref{sec:validation}.

\subsubsection{LCOGT}

The Las Cumbres Observatory Global Telescope \citep[LCOGT;][]{Brown:2013} 2.0\,m, 1.0\,m and 0.4\,m network nodes are located at Cerro Tololo Inter-American Observatory in Chile (CTIO), Siding Spring Observatory near Coonabarabran, Australia (SSO), South Africa Astronomical Observatory near Sutherland South Africa (SAAO), Teide Observatory on the island of Tenerife (TEID), McDonald Observatory near Fort Davis, TX, United States (McD), and Haleakala Observatory on Maui, Hawai'i (HAl). The MuSCAT3 multi-band imager \citep{Narita:2020} is installed on the LCOGT 2\,m Faulkes Telescope North at Haleakala Observatory. The image scale is $0\farcs27$ per pixel resulting in a $9.1\arcmin\times9.1\arcmin$ field of view. The 1\,m telescopes are located at all nodes except Haleakala and are equipped with $4096\times4096$ SINISTRO cameras having an image scale of $0\farcs389$ per pixel, resulting in a $26\arcmin\times26\arcmin$ field of view. The 0.4\,m telescopes are located at all nodes and are equipped with $2048\times3072$ pixel SBIG STX6303 cameras having an image scale of 0$\farcs$57 pixel$^{-1}$, resulting in a $19\arcmin\times29\arcmin$ field of view. All LCOGT images were calibrated by the standard LCOGT {\tt BANZAI} pipeline \citep{McCully:2018}, and differential photometric data were extracted using {\tt AstroImageJ} \citep{Collins:2017}.

\subsubsection{MuSCAT} \label{sssec:muscat}
The MuSCAT \citep[Multicolor Simultaneous Camera for studying Atmospheres of Transiting exoplanets;][]{Narita:2015} multi-color imager is installed at the 1.88\,m telescope of the National Astronomical Observatory of Japan (NAOJ) in Okayama, Japan. MuSCAT is equipped with three detectors for the Sloan $g'$, Sloan $i'$, and Sloan $z'_{s}$ band. The image scale is $0\farcs358$ per pixel resulting in a $6.1\arcmin\times6.1\arcmin$ field of view. MuSCAT data were extracted using the custom pipeline described in \cite{Fukui:2011}.

\subsubsection{MuSCAT2} \label{sssec:muscat2}
The MuSCAT2 multi-color imager \citep{Narita:2019} is installed at the 1.52~m Telescopio Carlos Sanchez (TCS) in the Teide Observatory, Spain. MuSCAT2 observes simultaneously in Sloan $g'$, Sloan $r'$, Sloan $i'$, and z-short. The image scale is $0\farcs44$ per pixel resulting in a in a $7.4\arcmin\times7.4\arcmin$ field of view. The photometry was carried out using standard aperture photometry calibration and reduction steps with a dedicated MuSCAT2 photometry pipeline, as described in \citet{Parviainen:2020}.

\subsubsection{MEarth-S} \label{sssec:mearth-s}
MEarth-South \citep{Irwin:2007} consists of eight 0.4\,m telescopes and observes from Cerro Tololo Inter-American Observatory, east of La Serena, Chile. Each telescope uses an Apogee U230 detector with a $29\arcmin\times29\arcmin$ field of view and an image scale of $0\farcs84$ per pixel. Results were extracted using the custom pipelines described in \cite{Irwin:2007}.

\subsubsection{El Sauce} \label{sssec:el_sauce}
The Evans 0.36\,m Planewave telescope is located at the El Sauce Observatory in Coquimbo Province, Chile. The telescope is equipped with a $1536\times1024$ pixel SBIG STT-1603-3 detector. The image scale is $1\farcs47$ per $2\times2$ binned pixel resulting in an $18.8\arcmin\times12.5\arcmin$ field of view.

\subsubsection{Deep Sky West} \label{sssec:deep_sky_west}
Deep Sky West is an Observatory in Rowe, NM. The 0.5\,m telescope is equipped with a Apogee U16M detector that has a image scale of $1\farcs09$ pixel$^{-1}$ resulting in a $37\arcmin\times37\arcmin$ field of view.

\subsubsection{Dragonfly}
The Dragonfly Telephoto Array is a remote telescope consisting of an array of small telephoto lenses roughly equivalent to a 1.0 m refractor housed at the New Mexico Skies telescope hosting facility, near Mayhill, NM, USA. Dragonfly uses SBIG STF8300M detectors that have an image scale of $2\farcs85$ pixel$^{-1}$, resulting in a $156\arcmin\times114\arcmin$ field of view. The data were reduced and analyzed with a custom differential aperture photometry pipeline designed for multi-camera image processing and analysis.

\subsubsection{SUTO-Otivar} \label{sssec:suto-ovitar}
The Silesian University of Technology Observatory (SUTO-Otivar) is an Observatory near Motril, Spain. The 0.3\,m telescope is equipped with a ZWO ASI 1600MM detector that has a image scale of $0\farcs685$~pixel$^{-1}$, resulting in a $18\arcmin\times13\arcmin$ field of view.

\subsubsection{Wellesley College Whitin Observatory} \label{sssec:whitin}
The Whitin observatory is a 0.7\,m telescope in Wellesley, MA. The $2048\times2048$ FLI ProLine PL23042 detector has an image scale $0\farcs68$~pixel$^{-1}$, resulting in a $23\farcm2\times23\farcm2$ field of view. 

\subsubsection{Adams Observatory} \label{sssec:adams_obs}
Adams Observatory is located at Austin College in Sherman, TX. The 0.6\,m telescope is equipped with a FLI Proline PL16803 detector that has a image scale of $0\farcs38$~pixel$^{-1}$, resulting in a $26\arcmin\times26\arcmin$ field of view.

\subsubsection{OAUV} \label{sssec:oauv-turia2}
The Observatori Astron\`{o}mic de la Universitat de Val\`{e}ncia (OAUV) is located near Valencia, Spain. The 0.3m telescope TURIA2 is equipped with a QHY 600 detector that has a image scale of 0\farcs68 pixel$^{-1}$, resulting in a 109\arcmin\ $\times$ 73\arcmin\ field of view.

\subsubsection{Lewin Observatory} \label{sssec:lewin_obs}
The Maury Lewin Astronomical Observatory is located in Glendora, CA. The 0.35\,m telescope is equipped with a SBIG STF8300M detector that has a image scale of $0\farcs84$ pixel$^{-1}$, resulting in a $23\arcmin\times17\arcmin$ field of view.

\subsubsection{ASP} \label{sssec:asp}
The Acton Sky Portal private observatory is in Acton, MA, USA. The 0.36\,m telescope is equipped with an SBIG Aluma CCD4710 camera having an image scale of $1\arcsec$ pixel$^{-1}$, resulting in a $17.1\arcmin\times17.1\arcmin$ field of view.

\subsubsection{WCO} \label{sssec:wco}
The Waffelow Creek Observatory (WCO) is located in Nacogdoches, TX. The 0.35\,m telescope is equipped with a SBIG STXL-6303E detector that has a image scale of $0\farcs66$ pixel$^{-1}$, resulting in a $34\arcmin\times23\arcmin$ field of view.

\subsubsection{PvDKO} \label{sssec:pvdko}
The Peter van de Kamp Observatory is located atop the Science Center at Swarthmore College in Swarthmore, PA. The 0.62\,m telescope has a QHY600 CMOS camera, which yields a $26\arcmin\times17\arcmin$ field of view. 

\subsubsection{TRAPPIST} \label{sssec:trappist}
The TRAnsiting Planets and PlanetesImals Small Telescope (TRAPPIST) North 0.6\,m telescope \citep{Barkaoui2019_TRAPPIST-North} is located at Oukaimeden Observatory in Morocco and TRAPPIST-South 0.6\,m telescope \citep{Gillon2011} is located at the ESO La Silla Observatory in Chile \citep{Jehin:2011}. TRAPPIST North is equipped with an Andor IKONL BEX2 DD camera that has an image scale of 0$\farcs$6 per pixel, resulting in a $20\arcmin\times20\arcmin$ field of view. TRAPPIST South is equipped with a FLI camera that has an image scale of 0$\farcs$64 per pixel, resulting in a $22\arcmin\times22\arcmin$ field of view. The image data were calibrated and photometric data were extracted using either {\tt AstroImageJ} or a dedicated pipeline that uses the {\tt prose} framework described in \citet{Garcia:2022}.

\subsubsection{ExTrA} \label{sssec:extra}
The Exoplanets in Transits and their Atmospheres (ExTrA) is sited at the ESO La Silla Observatory in Chile and consists of an array of three 0.6\,m telescopes. Image data were calibrated and photometric data were extracted using a custom pipeline described in \citet{Bonfils:2015}.

\subsubsection{SPECULOOS-S} \label{sssec:speculoos}
The SPECULOOS Southern Observatory consists of four 1\,m telescopes at the Paranal Observatory near Cerro Paranal, Chile. (Jehin et al. 2018). The telescopes are equipped with detectors that have an image scale of 0$\farcs$35 per pixel, resulting in a $12\arcmin\times12\arcmin$ field of view. The image data were calibrated and photometric data were extracted using a dedicated pipeline described in \citet{Sebastian:2020}.

\subsubsection{SAINT-EX} \label{sssec:saint-ex}
The SAINT-EX Observatory is located in San Pedro Mártir, Mexico. The 1.0\,m telescope is equipped with an Andor detector that has an image scale of 0$\farcs$34 per pixel, resulting in a $12\arcmin\times12\arcmin$ field of view. The image data were calibrated and photometric data were extracted using the SAINT-EX automatic reduction and photometry pipeline \citep[PRINCE;][]{Demory:2020}.

\subsubsection{CHAT} \label{sssec:chat}
The  0.7\,m Chilean-Hungarian Automated Telescope (CHAT) telescope is located at Las Campanas Observatory, in Atacama, Chile. Image calibration and photometric data were extracted using standard calibration and reduction steps and by a custom pipeline which implements bias, dark, and flat-field corrections.

\subsubsection{Observatori Astron\`{o}mic Albany\`{a}} \label{sssec:oaa}
The Observatori Astron\`{o}mic Albany\`{a} (OAA) is located in Albanyà, Girona Spain. The 0.4\,m telescope is equipped with a Moravian G4-9000 camera that has an image scale of $1\farcs44$ per $2\times2$ binned pixel resulting in a $36\arcmin\times36\arcmin$ field of view. 

\subsubsection{Lookout Observatory} \label{sssec:lookout_obs}
The Lookout Observatory is located in Colorado Springs, CO. The 0.5\,m telescope is equipped with a ZWO ASI1600MM Pro CMOS detector that has an image scale of $1\farcs46$ pixel$^{-1}$, resulting in a $152\arcmin\times101\arcmin$ field of view. The image data were calibrated and photometric data were extracted using the reduction and photometry pipeline described in \citet{Thomas:2021}. 

\subsubsection{Brierfield Private Observatory} \label{sssec:brierfield_obs}
The Brierfield Observatory is located near Bowral, N.S.W., Australia. The 0.36\,m telescope is equipped with a $4096\times4096$ Moravian 16803 camera with an image scale of 0$\farcs$74 pixel$^{-1}$, resulting in a $50\arcmin\times50\arcmin$ field of view.

\subsubsection{Caucasian Mountain Observatory} \label{sssec:CMO}
The Caucasian Mountain Observatory (CMO SAI MSU) houses a 0.6\,m telescope (RC600) and is located near Kislovodsk, Russia \citep{berdnikov2020v811ophperiod}. RC600 is equipped with an Andor iKon-L BV detector that has an image scale of $0\farcs67$~pixel$^{-1}$, resulting an a $22\arcmin\times22\arcmin$ field of view.

\subsubsection{Observatory de Ca l'Ou} \label{sssec:obs_de_calou}
 Observatori de Ca l'Ou (CALOU) is a private observatory in Sant Martí Sesgueioles, near Barcelona Spain. The 0.4\,m telescope is equipped with a $1024\times1024$ pixel FLI PL1001 camera having an image scale of $1\farcs$14 pixel$^{-1}$, resulting in a $21\arcmin\times21\arcmin$ field of view.
 
\subsubsection{Privat Observatory Herges-Hallenberg} \label{sssec:herges-hallenberg}
The Privat Observatory Herges-Hallenberg is a 0.28\,m telescope near Steinbach-Hallenberg, Germany. It is equipped with a Moravian Instrument G2-1600 detector that has an image scale of $1\farcs$02 pixel$^{-1}$, resulting an a $27\arcmin\times41\arcmin$ field of view. 

\subsubsection{Catania Astrophysical Observatory} \label{sssec:catania_obs}
The 0.91\,m telescope of the Catania Astrophysical Observatory is located on the slopes of Mt. Etna (1735\,m altitude) near Catania, Italy. The custom imaging camera uses as detector a 1024$\times$1024 KAF1001E CCD with an image scale of 0$\farcs$66 pixel$^{-1}$, resulting in a $11\farcm2\times11\farcm2$ field of view\footnote{\url{https://openaccess.inaf.it/handle/20.500.12386/764}}.

\subsubsection{Campo Catino Astronomical Observatory} \label{sssec:campo_catino}
The Campo Catino Astronomical Observatory (OACC) is located in Guarcino, Italy, and is equipped with a 0.8\,m RC telescope and a remote 0.6m CDK telescope located in El Sauce, Chile. In this work, iTelescope T17 was used, which is a 0.43\,m CDK telescope located at Siding Spring Observatory, equipped with a FLI  PL4710 CCD camera, providing a field of view of $15.5\arcmin\times15.5\arcmin$ and an image scale of $0\farcs92$ pixel$^{-1}$.

\subsubsection{RCO} \label{sssec:rco}
The 0.4\,m RCO telescope is located at the Grand-Pra Observatory in Valais Sion, Switzerland. The telescope is equipped with a FLI 4710 detector with an image scale of $0\farcs73$ pixel$^{-1}$, resulting in a $12.9\arcmin\times12.9\arcmin$ field of view. 

\subsubsection{CROW Observatory} \label{sssec:crow_obs}
The 0.36\,m telescope CROW Observatory is located in Portalegre, Portugal. It is equipped with a SBIG ST-10XME (KAF3200ME) detector that has an image scale of $0\farcs$66 pixel$^{-1}$, resulting an a $24\arcmin\times17\arcmin$ field of view. 

\subsubsection{MASTER-Ural} \label{sssec:master-ural}
The Kourovka observatory of Ural Federal University houses 0.4~m binocular MASTER-Ural telescope near Yekaterinburg, Russia. Each optical tube is equipped with an Apogee ALTA U16M detector with an image scale of 1$\farcs$85 pixel$^{-1}$, resulting an a 120$\arcmin~\times$~120$\arcmin$ field of view. The image data were calibrated, and photometric data were extracted using the reduction and photometry pipeline described in \citet{MASTER-Ural}.

\subsubsection{Kutztown University Observatory} \label{sssec:kutztown_obs}
The  0.6\,m telescope at Kutztown University Observatory is located near Kutztown, PA. The SBIG STXL-6303E detector has an image scale of $0\farcs76$ per $2\times2$ binned pixel, resulting in an $13\arcmin\times19.6\arcmin$ field of view. 

\subsubsection{Union College Observatory} \label{sssec:union_college_obs}
The Union College observatory houses a 0.51\,m telescope and is located in Schenectady, New York. The SBIG STXL detector has an image scale of $0\farcs93$ per $2\times2$ binned pixel, resulting in an $30\arcmin\times20\arcmin$ field of view. 

\subsubsection{George Mason University} \label{sssec:gmu}
The George Mason University 0.8\,m telescope near Fairfax, VA. The telescope is equipped with a $4096\times4096$ SBIG-16803 camera having an image scale of $0\farcs35$ pixel$^{-1}$, resulting in a $23\arcmin\times23\arcmin$ field of view. 

\subsubsection{Mt.\,Kent CDK700} \label{sssec:mko}
The University of Louisville's  MKO CDK700 telescope is located near Toowoomba, QLD, Australia at the University of Southern Queensland's Mt.\,Kent Observatory. It is a remotely operated Planewave Instruments 0.7-meter corrected Dall-Kirkham telescope with an Apogee U16M camera incorporating an OnSemi KAF-16803 CCD with $4096\times4096$ 9 $\mu\mathrm{m}$   0.41$^{\prime\prime}$ pixels and a $28^\prime \times 28^\prime$ field of view.

\subsubsection{Mt.\,Lemmon ULMT} \label{sssec:mt.lemmon}
The University of Louisville Manner telescope (ULMT) is located near Tucson, AZ USA at Mt.\,Lemmon Observatory. It is a remotely operated RC Optical Systems 0.61-meter Ritchie-Chr\'{e}tien telescope with a focal plane scale of  43~$^{\prime\prime}\mathrm{mm}$ with SBIG~STX~16803 and Apogee~U16M cameras incorporating OnSemi KAF-16803 CCDs with $4096\times4096$ 9 $\mu\mathrm{m}$ 0.39~$^{\prime\prime}$ pixels for a $27^\prime \times 27^\prime$ field of view .

\subsubsection{Mt.\,Stuart Observatory} \label{sssec:mt_stuart_obs}
The Mt.\,Stuart Observatory near Dunedin, New Zealand. The 0.32\,m telescope is equipped with a $3072\times2048$ SBIG STXL6303E camera with an image scale of 0$\farcs$88 pixel$^{-1}$ resulting in a $44\arcmin\times30\arcmin$ field of view.

\subsubsection{Fred L. Whipple Observatory} \label{sssec:FLWO}
The Fred Lawrence Whipple Observatory houses a 1.2\,m telescope and is located on Mt.\,Hopkins in Amado, AZ. The Fairchild CCD 486 detector has an image scale of $0\farcs672$ per $2\times2$ binned pixel, resulting in a $23\farcm1\times23\farcm1$ field of view. 

\subsubsection{Hazelwood Observatory} \label{sssec:hazelwood_obs}
The Hazelwood Observatory is located near Churchill, Victoria, Australia. The 0.32\,m telescope is equipped with a SBIG STT3200 camera with an image scale of 0$\farcs$55 pixel$^{-1}$, resulting in a $20\arcmin\times14\arcmin$ field of view.

\subsubsection{PEST} \label{sssec:pest}
The Perth Exoplanet Survey Telescope (PEST) is located near Perth, Australia. The 0.3 m telescope is equipped with a $5544\times3694$ QHY183M camera.  Images are binned 2x2 in software giving an image scale of 0$\farcs$7 pixel$^{-1}$ resulting in a $32\arcmin\times21\arcmin$ field of view. Prior to 23 March 2021 PEST was equipped with a $1530\times1020$ SBIG ST-8XME camera with an image scale of 1$\farcs$2 pixel$^{-1}$ resulting in a $31\arcmin\times21\arcmin$ field of view.  A custom pipeline based on {\tt C-Munipack}\footnote{\url{http://c-munipack.sourceforge.net}} was used to calibrate the images and extract the differential photometry.

\subsubsection{Salerno University Observatory} \label{sssec:salerno_obs}
The Salerno University Observatory houses a 0.6\,m telescope and is located in Fisciano, Italy. The telescope is equipped with a FingerLakes Instrument Proline L230 that has a $21\arcmin\times21\arcmin$ field of view with $0\farcs61$ pixel$^{-1}$.

\subsubsection{Villa '39} \label{sssec:villa39}
The Villa '39 Observatory is located in Landers, CA. The 0.35\,m telescope is equipped  with KAF16803 detector that has an image scale of $0\farcs94$ pixel$^{-1}$ resulting in a $32.5\arcmin\times32.5\arcmin$ field of view.

\subsubsection{Solaris SLR2} \label{sssec:solaris_slr2}

The SLR2 is one of four automated telescopes of the {\it Solaris} network, owned and operated by the N. Copernicus Astronomical Center of the Polish Academy of Sciences. SLR2 is a 0.5-m telescope located in SAAO, equipped an Andor Ikon-L camera having an image scale of $0\farcs367$ pixel$^{-1}$, resulting in a $12\arcmin\times12\arcmin$ field of view.

\subsubsection{Wild Boar Remote Observatory} \label{sssec:wild_boar_obs}

The Wild Boar Remote Observatory is a private observatory located in San Casciano in val di Pesa (Firenze), Italy. It has a remotely-operated 0.23\,m Schmidt-Cassegrain telescope equipped with an Sbig ST-8 XME CCD.

\subsubsection{Gruppo Astrofili Catanesi} \label{sssec:gruppo_astrofili_catanesi}

The Gruppo Astrofili Catanesi is a private observatory located in Catania, Italy. It possesses a 0.25\,m Newtonian telescope with an Sbig ST-7 XME CCD.

\subsubsection{Ground Survey and Space Data}
We used archival ground-based survey data and related follow-up observations from HATSouth \citep{Bakos:2013} and WASP \citep{Pollacco:2006} that pre-dated the TESS mission to help disposition some of the planet candidates. We also used results from the Gaia-TESS collaboration \citep{Panahi:2022}, which is a joint analysis of TESS photometry and unpublished Gaia time-series photometry, to disposition some planet candidates. Additionally, we used archival data taken by \ac{ZTF} for a subset of the best-in-class \acp{TOI} to determine if their signals were on-target. To accomplish this, we implemented the code DEATHSTAR \citep{deathstar} which is further described in Section \ref{ssec:phot_vet}.

\subsection{Reconnaissance Spectroscopy} \label{ssec:sg2_spectroscopy}

\ac{TFOP}'s SG2 performed ground-based reconnaissance spectroscopy on a subset of targets in our best-in-class samples. These observations are crucial to constraining the mass of potential stellar or planetary companions to the host star and for refining the stellar parameters to be used in future analysis. Below we detail the observatories, instruments, and data reduction methods used to obtain the reconnaissance spectroscopy used in our analysis. See Section \ref{ssec:spec_vet} for further discussion on how reconnaissance spectroscopy is used in our vetting procedures.

\subsubsection{TRES} \label{sssec:tres}

Reconnaissance spectra were obtained with the Tillinghast Reflector Echelle Spectrograph \citep[TRES;][]{gaborthesis} which is mounted on the 1.5m Tillinghast Reflector telescope at the Fred Lawrence Whipple Observatory (FLWO) located on Mount Hopkins in Arizona. TRES is a fiber-fed echelle spectrograph with a wavelength range of 390-910nm and a resolving power of $R\sim$44,000. Typically, 2-3 spectra of each target are obtained at opposite orbital quadratures to check for large velocity variation due to a stellar companion. The spectra are also visually inspected to ensure a single-lined spectrum. The TRES spectra are extracted as described in \cite{buchhave2010} and stellar parameters are derived using the Stellar Parameter Classification tool \citep[SPC;][]{buchhave2012}. SPC cross correlates an observed spectrum against a grid of synthetic spectra based on Kurucz atmospheric models \citep{kurucz1992} to derive effective temperature, surface gravity, metallicity, and rotational velocity of the star.

\subsubsection{FIES} \label{sssec:fies}

We used the FIbre-fed Echelle Spectrograph \citep[FIES;][]{telting2014}, a cross-dispersed high-resolution spectrograph mounted on the 2.56\,m Nordic Optical Telescope \citep[NOT;][]{djupvik2010}, at the Observatorio del Roque de los Muchachos in La Palma, Spain. FIES has a maximum resolving power of $R\sim67,000$, and a spectral coverage that ranges from 3760~\AA\ to 8820~\AA. The data were extracted as described in \citet{buchhave2010}.

\subsubsection{CHIRON} \label{sssec:chiron}

We obtained high resolution spectroscopic vetting observations with the CHIRON spectrograph for a number of the TESS planet candidates. CHIRON is a high resolution echelle spectrograph on the SMARTS 1.5\,m telescope at the Cerro Tololo Inter-American Observatory, Chile \citep{2013PASP..125.1336T}. We typically make use of the spectrograph in its `slicer' mode, fed via a fiber through an image slicer to achieve a spectral resolving power of $R\sim80,000$ over the wavelength range of $4100-8700\,\AA{}$. Spectral extraction is performed via the official CHIRON pipeline \citep{2021AJ....162..176P}. We derive radial velocities and spectral line profiles via a least-squares deconvolution \citep{1997MNRAS.291..658D} between the observed spectra and a non-rotating synthetic spectral template that matches the atmospheric parameters of the target star. Radial and line broadening velocities are derived by modeling the line profile as per \citet{2020ApJ...892L..21Z}. For some of the faintest host stars ($V\gtrsim12.5$), we use CHIRON in `fiber' mode, which achieves a lower resolving power of $R\sim28,000$, but yields similar vetting information at lower precision.

\subsubsection{Keck/HIRES} \label{sssec:hires}

We obtained radial velocity data using the Keck Observatory HIRES spectrometer \citep{vogt1994proc} on the Keck I telescope atop Mauna Kea. We use the iodine cell technique pioneered by \cite{butler1996attaining}. Radial velocities were measured using an iodine gaseous absorption cell as a precision velocity reference, placed just ahead of the spectrometer slit in the converging beam from the telescope. Doppler shifts from the spectra are determined with the spectral synthesis technique described by \cite{butler1996attaining}. For this velocity analysis, the iodine region of the echelle spectrum was subdivided into $\sim$700 wavelength chunks of 2 \r{A} each. Each chunk provided an independent measure of the wavelength, PSF, and Doppler shift. The final measured velocity is the weighted mean of the velocities of the individual chunks.

\subsubsection{HARPS-N} \label{sssec:harps-n}

HARPS-N is a fiber-fed, cross-dispersed echelle spectrograph with a spectral resolution of 115,000 mounted at the 3.58\,m Telescopio Nazionale Galileo (TNG) in La Palma island, Spain. It covers the visible wavelength range from 3830 to 6900 \AA \citep{Cosentino2012}. Spectra extraction and reduction was carried out using the HARPS-N data reduction software (DRS). Radial velocities were obtained by cross-correlating the spectra with a numerical mask close to the stellar spectral type (e.g., \citealt{Pepe2002hd108147}).

\subsubsection{PFS} \label{sssec:pfs}

The Planet Finder Spectrograph \citep[PFS;][]{crane2006carnegie,crane2008carnegie,crane2010carnegie} is installed at the 6.5 m Magellan/Clay telescope at Las Campanas Observatory. Targets were observed with the iodine gas absorption cell of the instrument, adopting an exposure time of 1200 s and using a 3 × 3 CCD binning mode to minimize read noise. Targets were also observed without the iodine cell in order to generate the template for computing the RVs, which were derived following the methodology of \cite{butler1996attaining}.

\subsubsection{CORALIE} \label{sssec:coralie}

The CORALIE high-resolution echelle spectrograph is mounted on the Swiss Euler 1.2\,m telescope at La Silla Observatory, Chile \citep{CORALIE}. The spectrograph is fed by a 2'' on-sky science fibre and a secondary B-fibre which can be used for simultaneous wavelength calibrations with a Fabry-Perot etalon or pointed on-sky to monitor background contamination. CORALIE has a spectral resolution of $R\sim$60,000 and reaches an RV precision of 3 m/s when photon-limited. Stellar RV measurements are extracted via cross-correlation with a mask \citep{baranne96, pepe2002}, using the standard CORALIE data-reduction pipeline. TOIs are vetted using several CCF line-diagnostics such as bisector-span, FWHM. We also check for mask-dependent RVs, SB2, SB1 and visual binaries. False positives are routinely reported to EXOFOP-TESS and data made available through the DACE platform\footnote{e.g. \url{https://dace.unige.ch/radialVelocities/?pattern=TOI-128}}.

\subsubsection{Minerva-Australis} \label{sssec:minerva-australis}

We carried out spectroscopic observations using the MINERVA-Australis facility \citep{addison2019minerva}. MINERVA-Australis consists of an array of four independently operated 0.7 m CDK700 telescopes situated at the Mount Kent Observatory in Queensland, Australia. Each telescope simultaneously feeds stellar light via fiber optic cables to a single KiwiSpec R4-100 high-resolution (R = 80,000) spectrograph \citep{barnes2012kiwispec} with wavelength coverage from 480 to 620 nm. Radial velocities for the observations are derived for each telescope by cross-correlation, where the template being matched is the mean spectrum of each telescope. The instrumental variations are corrected by using simultaneous ThAr arc lamp observations.

\subsubsection{NRES} \label{sssec:nres}

The Network of Robotic Echelle Spectrographs \citep[NRES;][]{Siverd18} is a set of four identical fiber-fed spectrographs on the 1m telescopes of LCOGT \citep{Brown:2013}. The NRES units are located at the LCOGT nodes at Cerro Tololo Inter-American Observatory, Chile; McDonald Observatory, Texas, USA; South African Astronomical Observatory, South Africa; and Wise Observatory, Israel. The spectrographs deliver a resolving power of $R\sim$53,000 over the wavelength range 3800-8600 \AA. The data were reduced and radial velocities measured using the \texttt{BANZAI-NRES} pipeline \citep{McCully22}. We measured stellar parameters from the spectra using a custom implementation of the \texttt{SpecMatch-Synth} package\footnote{\url{https://github.com/petigura/specmatch-syn}} \citep{petigura2017cks1}.

\subsubsection{FEROS} \label{sssec:feros}
The Fiber-fed Extended Range Optical Spectrograph \citep[FEROS;][]{kaufer:1998} spectrograph is a high resolution (R$\sim$48,000) echelle spectrograph installed at the MPG2.2m telescope at the ESO La Silla Observatory, Chile. FEROS covers the spectral range between 350 and 920 nm and has a comparison fiber to trace instrumental radial velocity drifts during the science exposures with a thorium argon lamp. FEROS data are processed with the automated \texttt{ceres} pipeline \citep{Brahm:2017} that generates precision radial velocities and bisector span measurements starting from the raw images which are reduced, optimally extracted and wavelength calibrated before cross-correlating the spectrum with a G2-type binary mask.

\subsection{High-resolution Imaging} \label{ssec:sg3_imaging} 

As part of our standard process for validating transiting exoplanets to assess the possible contamination of bound or unbound companions on the derived planetary radii \citep{ciardi2015}, we also observed a subset of the unconfirmed \acp{TOI} in our best-in-class sample with a combination of near-infrared adaptive optics (AO) imaging and optical speckle interferometry at a variety of observatories including Gemini, Keck, Lick, Palomar, VLT, and WIYN Observatories. The combination of the observations in multiple filters enables better characterization for any companions that may be detected and improves the sensitivity to different types of false positive scenarios (e.g. bound low-mass companions, background stars, etc.). See Sections \ref{ssec:companions} and \ref{sec:validation} for further discussion on how high-resolution imaging was incorporated into our vetting and validation analyses, respectively.

\subsubsection{Near-Infrared AO Imaging}  Near-infrared AO observations are performed with a dither pattern to enable the creation of a sky-frame from a median of the science frames. All science frames are flat-fielded (which are dark-subtracted) and sky-subtracted.  The reduced science frames are combined into a single combined image using an intra-pixel interpolation that conserves flux, shifting the individual dithered frames by the appropriate fractional pixels; the final resolution of the combined dithers was determined from the full-width half-maximum of the point spread function. The sensitivities of the final combined AO images were determined by injecting simulated sources azimuthally around the primary target every $20^\circ $ at separations of integer multiples of the central source's FWHM \citep{furlan2017}. The brightness of each injected source was scaled until standard aperture photometry detected it with $5\sigma $ significance.  The final $5\sigma $ limit at each separation was determined from the average of all of the determined limits at that separation and the uncertainty on the limit was set by the rms dispersion of the azimuthal slices at a given radial distance.

\subsubsection{Optical Speckle Imaging}

High-Resolution optical speckle interferometry was performed using the `Alopeke and Zorro instruments mounted on the Gemini North and South telescopes respectively \citep{scott2021geminiimagers,howell2021nasa}. These identical instruments provide simultaneous speckle imaging in two bands (562\,nm and 832\,nm) with output data products including a reconstructed image and robust contrast limits on companion detections \citep{howell2011specklekepler}. For each observed source, the final reduced data products contain 5$\sigma$ contrast curves as a function of angular separation, information on any detected stellar companions within the angular range of $\sim0.03\arcsec$ to $1.2\arcsec$ (delta magnitude, separation, and position angle), and reconstructed speckle images in each band-pass. The angular separation sampled, from the 8\,m telescope diffraction limit (20 mas) out to $1.2\arcsec$, can be used to set spatial limits in which companions were or were not detected.

\section{Vetting} \label{sec:vetting}

In order to determine the planetary nature of each target, we performed a uniform vetting procedure on each of the unconfirmed candidates. This included utilizing a mix of publicly-available resources and follow-up observations obtained by \ac{TFOP}. We outline our overall procedure in schematic form in the middle panel of Figure \ref{fig:vetting_procedure}. We ran each target through as many steps of our vetting procedure as possible given the availability of resources at the time of analysis in early 2023 since not all targets had the resources to complete each step in our procedure. 

Although our vetting procedure checked for a number of false positive indicators, we refrained from classifying a target as a likely false positive unless multiple false positive indicators suggested that the origin of the transit signal could not have been a planet. Our conservative approach to vetting passed most targets on to statistical validation and provided invaluable information to be used in conjunction with the results from our validation analysis to make a final determination, such as if the signal is on-target and if there were any potentially conatimating sources contained in the light curve's extraction aperture. In this way, vetting served as a complement to a more holistic determination of the planetary nature that accounts for a larger number of factors than any individual analysis alone could provide.

For all of our vetting, we used the orbital and planetary parameter values posted on ExoFOP\footnote{\url{https://exofop.ipac.caltech.edu/tess/}} unless follow-up observations revealed more accurate or precise values for a given parameter, in which case, the parameters obtained from follow-up were used. There were seven targets with ambiguous periods contained in our best-in-class samples from the query of the Exoplanet Archive (TOIs 706.01, 1856.01, 1895.01, 2299.01, 4317.01, 5575.01, and 5746.01). These targets were single transits that transited again in one or more later \ac{TESS} sectors, but without measuring two or more consecutive transits, the period could not be confidently determined. The orbital period of TOI-5575.01 was was uniquely determined to be 32.07 d through follow-up observations over the course of our analysis\footnote{\url{https://exofop.ipac.caltech.edu/tess/target.php?id=160162137}}. We propagated this updated period throughout our analysis and report the planet's updated parameters in our final best-in-class sample. Since we are unable to obtain the true periods of these remaining six targets without a concerted observing campaign, we analyzed them with the reported ExoFOP periods that represent upper limits to the true periods. Shorter periods would likely result in higher equilibrium temperatures which, although this would boost the \ac{TSM}, could place these targets in a different temperature bin where they may not rank in the top 5 targets in their planetary radius bin. We recognize that the periods and therefore amenability to atmospheric characterization with \ac{JWST} may change for these targets, but we include them in our best-in-class samples to emphasize their potential as prime \ac{JWST} targets and encourage their further study.

\subsection{\textit{TESS} SPOC Data Validation Report} \label{ssec:dv_reports}

$\sim$92$\%$ of the targets in our best-in-class samples were either discovered by (or at the very least run through) the \ac{TESS} \ac{SPOC} pipeline \citep{jenkins2016spoc} at NASA's Ames Research Center. This SPOC pipeline performs a number of tasks on each target including light curve extraction to generate Simple Aperture Photometry (SAP) light curves \citep{twicken:PA2010SPIE,morris:PA2020KDPH} and systematic error correction to generate \ac{PDCSAP} light curves. The pipeline also searches for potential planets as well as performs a suite of diagnostic tests in the Data Validation (DV) module to help adjudicate the planetary nature of each signal \citep{twicken2018DVdiagnostics, Li2019DVmodelFit}. Upon running the pipeline, the outputs were reviewed by the \ac{TESS} \ac{TOI} Working Group (TOI WG) to perform initial vetting. This initial vetting has already been performed by the TOI WG for all of our targets, but we reviewed the \ac{SPOC} pipeline outputs again to ensure nothing was missed.

The DV module includes a depth test of the odd and even transits, a statistical bootstrap test that accounts for the non-white nature of the observation noise to estimate the probability of a false alarm from random noise fluctuations, a ghost diagnostic test to compare the detection statistic of the optimal aperture against that of a halo with a 1 pixel buffer around the optimal aperture, and a difference image centroid test. At the conclusion of these tests, the module synthesizes a summary of the results for each individual test, including assigning a pass/fail disposition for each test. We used the results of each of these tests in our vetting efforts to help determine if a target was a likely planet, likely false positive, or false alarm.

In addition to the DV module results, we also determined if the period was ambiguous for a given target due to nonconsecutive transits from gaps in the \ac{TESS} data. Although not a false positive indicator, this was flagged for future reference in downstream analyses. We also checked the light curves for significant photometric modulation indicative of stellar activity that could pose a problem in future vetting and validation analysis. In the absence of \ac{SPOC} DV results, we still inspected the light curve and ephemerides for an ambiguous period or photometric modulation using available, published light curves such as those from MIT's \ac{QLP}.

\subsection{DAVE Vetting from Cacciapuoti, et al.} \label{ssec:tt9}

A subset ($\sim$15$\%$) of our targets had already been vetted not only by the \ac{TESS} TOI WG, but by an independent team using \ac{DAVE}. The results of this vetting were collated in \cite{cacciapuoti2022tt9} where each of the 999 targets vetted were assigned a final disposition as to the target's planetary nature.

\ac{DAVE} is an automated vetting pipeline built upon many of the tools developed for vetting planets in \textit{Kepler} data (e.g. RoboVetter, \citealt{coughlin2014robovetter}) and has been used extensively in vetting planets for \ac{TESS} \citep[e.g.][]{gilbert2020toi700, hord2021hotjupiters, quintana2023toi2095}. \ac{DAVE} performs two sets of vetting tests: 1) light curve-based vetting tests searching for odd/even transit depth differences, secondary eclipses, and light curve modulations and 2) image-based centroid tests to check the photometric motion on the \ac{TESS} image during transit.

For the targets in our best-in-class samples that were also contained in the \cite{cacciapuoti2022tt9} catalog, we included their dispositions in our vetting analysis. Since there is overlap between the tests performed by the \ac{TESS} \ac{SPOC} pipeline and \ac{DAVE}, we treat the two as independent checks of one another and review the results in comparison.

\subsection{Reconnaissance Photometry} \label{ssec:phot_vet}

Due to the large 21'' pixel size of \ac{TESS}, ground-based photometry at higher spatial resolutions is crucial in determining whether a transit-like feature is occurring on-target or is the result of a background target in the starfield that may have been blended within the \ac{TESS} pixel. Stars nearby the target are checked for deep EBs that could cause the observed transits and are ruled out on a case-by-case basis. Any deviations from an on-time transit are also noted. These often occur due to uncertainties in the period or mid-transit time reported by ExoFOP but may be caused by gravitational interactions within the system. If the period deviates significantly from the reported period, the ephemerides are refined based on the ground-based photometric observations. This was the case for multiple targets, especially those with fewer sectors of \ac{TESS} data or those with an ambiguous period.

In addition to checking which star the transit-like feature originates from, ground-based photometry uses multiple filters to check for possible chromaticity in the transit depth that would indicate an eclipsing binary rather than a planet is causing the transit. A light curve is also extracted from the target star with a small aperture to mitigate the contamination from nearby stars. The transit depth is measured to ensure that it is not only consistent across wavelength bandpasses, but is the right depth to cause the transit observed in the \ac{TESS} data.

\ac{TFOP}'s SG1 synthesizes the results of the photometric observations for each target into a single disposition describing the confidence with which a signal can be considered on-target. We utilized these dispositions and observations when determining which background stars to consider as potential sources of astrophysical false positives in our vetting analysis.

In addition to the photometry gathered by SG1, we also utilized the code \texttt{DEATHSTAR} \citep{deathstar} to search archival images from \ac{ZTF} for the transit signal. 

DEATHSTAR attempts to either confirm or refute exoplanet detections with already available ground-based data from \ac{ZTF} by extracting light curves for each star in a 2.5 arcminute field and plotting them for manual verification of the actual signal location. In this way we can often tell if an unconfirmed \ac{TOI} is an exoplanet transiting in front of the target star or an eclipsing binary on a nearby fainter star. 
DEATHSTAR creates plots for each extracted light curve and displays them in custom sheets for us to easily find the source of the transiting signal. We work with SG1 in checking these results with the SG1 Observation Coordinator sheet and sending them to reduce extraneous telescope follow-up time. For deeper transit depths on-target (ranging from 1-3\%), DEATHSTAR has been able to confirm on-target detections. Because the target stars are bright ($J$ < 13 mag) and given ZTF's sensitivity, we were able to check for and rule out eclipsing binaries among the surrounding stars in the TESS apertures down to the faintest stars that could account for the transit depth. Due to ZTF’s multiple filters ($g$, $r$, and $i$ bands), we can constrain the chromaticity of the transit signal, which can also indicate or help rule out false positives. In most of the cases for these targets, the depth was much shallower than a percent, rendering the transits undetectable by DEATHSTAR on-target, but we still cleared all the surrounding stars in the field for being potential eclipsing binaries, showing the transit signal must originate from the target by process of elimination.

\subsection{Reconnaissance Spectroscopy} \label{ssec:spec_vet}

Although only a subset of the targets in our sample had ground-based spectroscopic observations available, these data provided strong constraints on the presence of bound companions in the target system that photometry is unable to capture. Spectroscopy alone is often able to determine if the stellar spectrum is composite which would indicate the presence of a bound stellar mass companion. The presence of a composite spectrum with orbital motion that is consistent with the \ac{TESS} ephemeris was an automatic likely false positive designation for the targets in our samples but only applied to one target (TOI-4506.01).

For most targets, two spectroscopic observations were taken at opposite quadratures assuming a circular orbit at the photometric ephemeris and compared to the photometric ephemeris to determine if they were in-phase. Spectroscopic data at opposite quadratures that are out of phase with the photometric ephemeris could indicate the presence of a large stellar-mass object instead of a planet, although this could also indicate a long-term trend in the system due to additional bodies in the system or an eccentric orbit rather than a false positive scenario. For reconnaissance spectroscopy that was in-phase with the photometric ephemeris, the semi-amplitude of the measurements at quadrature was used to constrain the mass of the object producing the transit signal, potentially ruling out stellar masses and providing evidence for the planetary nature of the body.

By virtue of modeling the stellar spectrum, reconnaissance spectroscopy also has the potential to measure parameters such as the effective temperature, metallicity, and $v$ sin$i$ of the host star. Where possible, we used these measured values rather than those from \ac{TIC} or \textit{Gaia} DR3.

Similar to SG1, \ac{TFOP}'s SG2 also synthesizes reconnaissance spectroscopic observations into a disposition for each target. These dispositions capture the confidence that the target is a planetary mass object and is suitable for precision radial velocity observations to determine the orbit and constrain the mass further. We broadly utilized these dispositions when vetting to determine whether a target can be safely deemed a likely false positive or should continue to statistical validation analysis. There were multiple cases where reconnaissance spectroscopy existed but the stellar activity or rotational broadening of spectral features precluded anything but upper limits on the masses of potential companions.

\subsection{Imaging Constraints} \label{ssec:companions}

As a complement to ground-based photometry and reconnaissance spectroscopy, high-resolution imaging can provide strict constraints on the presence of stellar companions in the system or nearby background targets that could potentially contaminate the target signal. Each target was first cross-referenced with the \textit{Gaia} DR3 catalogue to determine if there are any resolved nearby stars within a few arcseconds of the target star. In a handful of cases, \textit{Gaia} resolved nearby stars at similar parallaxes to targets in our best-in-class samples. While not a definite indicator of a false positive, the presence of a nearby companion at a similar parallax invited further scrutiny for that particular target. In those cases, we cross-referenced the nearby star with other follow-up observations where possible to determine if the star observed by \textit{Gaia} may be the cause of anomalies and potential false positive indicators in the ground-based photometry or reconnaissance spectroscopy.

We also utilized speckle or adaptive optics (AO) imaging available on ExoFOP (see Section \ref{ssec:sg3_imaging}) that observed each planet candidate in a more targeted manner at a higher angular resolution than \textit{Gaia}. These observations allowed us to search for bound companions or background stars that may contaminate the photometry or cause the observed transit signal. These observations were also cross-referenced with other follow-up observations to determine how strongly false positive or dilution scenarios can be constrained or if the signal is likely not due to a planet. The sensitivity curves that these observations produced were also used in our statistical validation analysis (Section \ref{sec:validation}).

\section{Validation} \label{sec:validation}

While vetting is an integral step in determining whether a periodic signal is indeed due to the presence of a planet, it cannot alone demonstrate that a signal is not a false positive. The preferred method for determining whether a signal is a planet is a mass measurement through radial velocity (RV) observations, however these oftentimes require a significant commitment of resources and time on targets that may not prove to be planets. 

In lieu of a mass measurement, statistics can be used to validate the target rather than confirm it. Statistical validation of a target often only requires photometric and imaging observations as well as planetary and orbital parameters input into one or multiple statistical validation software packages. Targets that are validated to a greater than 99$\%$ confidence threshold are considered planets despite not having a mass measurement \citep{morton2012vespa,giacalone2021triceratopsvetting}. Since the time and observational resources required to validate a planet are far less than required to obtain a mass measurement, statistical validation serves as an excellent intermediate step to weed out targets that are very likely not planets in order to better streamline and prioritize the RV observations required to confirm a target as a bona fide planet.

In the case of our best-in-class samples, since there are undoubtedly false positives among the unconfirmed planet candidates, we performed statistical validation on all candidate planets to not only determine which targets are most likely to be true planets, but which merit follow-up with RV observations. To do this, we run the statistical validation software \vespa \citep{morton2012vespa, morton2015vespa} and \triceratops \citep{giacalone2020triceratops} on each of our unconfirmed targets in both the transmission and emission spectroscopy samples.

For all of our targets, we use the orbital and planetary parameters from ExoFOP unless the follow-up observations reported refined parameters (Section \ref{sec:observations}), in which case the refined parameters were used. For \vespa, this also included stellar parameters. \ac{TESS} photometry was used to produce the phase-folded transits used in both \vespa and \triceratops. When possible, we favored light curves produced by the \ac{TESS} \ac{SPOC} at the shortest cadence available since shorter cadence \ac{TESS} data have been shown to be more photometrically precise when binned than data taken at the binned cadence itself \citep{huber2022tessasteroseismology}. A small subset of targets did not have \ac{SPOC} \ac{PDCSAP} light curves, in which case we used light curves produced by MIT's \ac{QLP}.

\subsection{\vespa} \label{ssec:vespa}

\vespa \citep{morton2012vespa, morton2015vespa} was originally developed for use on \textit{Kepler} data and compares the input orbital and planetary parameters as well as the phase-folded transit against a number of astrophysical false positive scenarios to determine the likelihood that the signal can be produced by each false positive population. Currently, \vespa tests against the hypotheses that the signal is a blended background or foreground EB (BEB), the target itself is an EB, or the target is a hierarchical-triple system where two of the components form an EB (HEB). To do this, \vespa simulates a representative population of each false positive scenario at the observed period and calculates the priors of each scenario, accounting for the probability that the scenario is contained within the photometric aperture, the probability of an orbital alignment that would cause an observable eclipse, and the probability that the eclipse could mimic a transit. A TRILEGAL simulation \citep{girardi2005startrilegal, girardi2012trilegal} is used to simulate the background starfield for each target when calculating the priors. The likelihoods of each scenario are then calculated by modeling the shape of the eclipse for each instance of each false positive population and fitting it to the observed light curve. The priors and likelihoods are finally combined to calculate the total \ac{FPP} of the input transit signal. Signals with an \ac{FPP} $<$ 0.01 are considered statistically validated.

Beyond the phase-folded light curve and planetary and orbital parameters, \vespa can also intake sensitivity curves from high-resolution imaging to rule out portions of the false positive parameter space. Additionally, \vespa takes the maximum photometric aperture radius as an input to use in calculations of the BEB prior. We set this parameter to $42\arcsec$, the size of two \ac{TESS} pixels. This is very conservative since the difference image centroiding results from the \ac{SPOC} DV analysis often constrain the location of the target star to within a fraction of a pixel of the location of the source of the transit.

\vespa assumes that the signal originates on-target, which we have attempted to show for as many targets in our sample as possible (see Section \ref{sec:vetting}). We urge caution in the interpretation of the results from \vespa in the cases where the signal was not demonstrated to be on-target.

\subsection{\triceratops} \label{ssec:triceratops}

\begin{figure*}
    \centering
    \includegraphics[width=0.99\textwidth]{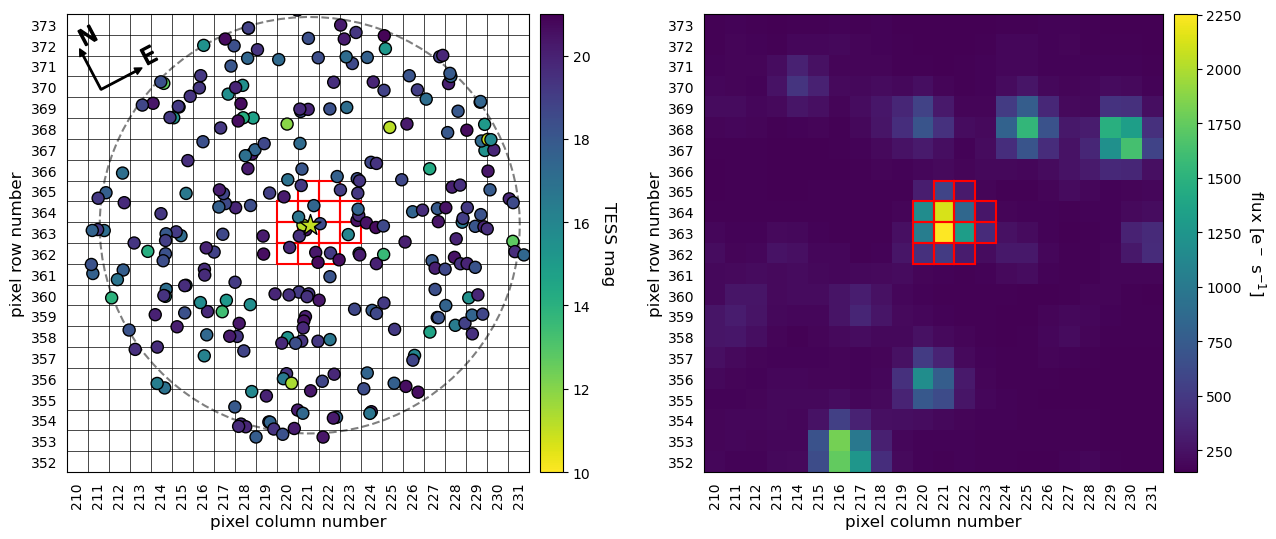}
    \caption{Starfield around TOI-4336.01 in \ac{TESS} Sector 38 used by \triceratops in its \ac{FPP} and \ac{NFPP} calculations. \textit{Left:} plot of the positions of each star within 2.5 arcminutes centered on the target with the color of each point representing the \ac{TESS} magnitude of the star. The overlaid grid denotes the \ac{TESS} pixel borders with pixel column and row numbers labeled on the X and Y axis, respectively. The dashed gray circle represents a distance of 2.5 arcminutes and the red squares denote the extraction aperture used by the \ac{SPOC} when generating the PDC$\_$SAP light curve for this \ac{TESS} sector. \textit{Right:} Same as left but instead of displaying each background star near the target, \ac{TESS} data are shown. The \ac{SPOC} extraction aperture is in red and the colormap represents the flux captured by each \ac{TESS} pixel.}
    \label{fig:triceratops_starfield}
\end{figure*}

Similar to \vespa, \triceratops \citep{giacalone2020triceratops} compares the user-provided phase-folded transit, orbital, and stellar parameters against a set of astrophysical false positive scenarios to rule out portions of parameter space in which the false positive scenarios can remain viable. The methodology of \triceratops is identical to \vespa in many respects, however, in contrast to \vespa, \triceratops was developed specifically for \ac{TESS} and accounts for the real sky background of each target out to 2.5' as well as the \ac{TESS} point spread function and aperture used to extract the photometric light curve in each sector of \ac{TESS} data. An example of what \triceratops considers in this portion of its analysis is seen in Figure \ref{fig:triceratops_starfield}.

For each target, we used the extraction apertures produced by the \ac{TESS} \ac{SPOC} contained within the headers of the \ac{SPOC} PDC-SAP light curves queried by \texttt{lightkurve} \citep{lightkurve2018lightkurve} on a sector-by-sector basis. For the targets missing \ac{SPOC} PDC-SAP light curves from some or all \ac{TESS} sectors they were observed in, we used a standard aperture of 5$\times$5 \ac{TESS} pixels. This is larger than any of the PDC-SAP apertures and is the \triceratops default for sectors without provided apertures.

When accounting for nearby background stars for each target, \triceratops queries the \ac{TIC}v8 for the stellar parameters of each star. The \ac{TIC} is based heavily on the \textit{Gaia} DR2 data release, which has since been updated by \textit{Gaia} DR3. Therefore, in our analysis, we queried the RA, Dec, mass, effective temperature, parallax, and \textit{Gaia} \textit{G} magnitude of the host star \textit{Gaia} DR3 Catalog for use in our analysis in lieu of using the values provided by the \ac{TIC}. To convert the \textit{Gaia} magnitude to \ac{TESS} magnitude, we used Equation 1 from \cite{stassun2019tic} which is valid for dwarfs, subgiants, and giants of any metallicity. We then cross-referenced each \textit{Gaia} target with the 2MASS catalog \citep{skrutskie20062mass} to obtain J, H, and K magnitudes where available as these magnitudes are used by \triceratops in its estimation of false positive probability.

Additionally, we included follow-up constraints into our analysis with \triceratops. When available, we included a contrast curve from high-resolution imaging to constrain the existence of additional stellar mass companions in the system. Unlike \vespa, \triceratops accepts only a single contrast curve per target, so in the case a target possessed multiple contrast curves from follow-up observations, we included only the contrast curve that provided the greatest imaging contrast magnitude agnostic of bandpass to most stringently constrain possible companions in the system. Furthermore, our photometric follow-up allowed us to clear individual nearby stars of potentially harboring EBs that would cause the observed transit signal on target. Background stars that were definitively determined to not be EBs at the target period or have an eclipse depth that could cause the observed transit on-target were discarded from consideration as potential sources of a false positive. Targets whose transits were observed on-target had all background stars removed from false positive consideration. As recommended by \cite{giacalone2021triceratopsvetting}, we ran multiple trials of the \triceratops \ac{FPP} calculation for each target, with a minimum of 10 trials per target and report the mean of these \acp{FPP}.

\triceratops provides not only a final \ac{FPP} value, but also a \ac{NFPP} value that encapsulates the probability that the signal originates from a star other than the target. \cite{giacalone2021triceratopsvetting} defines validated planets as signals with \ac{FPP} $<$ 0.015 and \ac{NFPP} $<$ 10$^{-3}$ and outlines a separate category for marginal validations when \ac{FPP} $<$ 0.5 and \ac{NFPP} $<$ 10$^{-3}$. We adopt these categories in our determination of the planetary nature for our best-in-class samples. We extend the marginal validations category to \vespa, which does not explicitly have such a distinction. In the case of \vespa, we conservatively set the marginal validation threshold to \ac{FPP} $<$ 0.25, lower than that of \triceratops.

\cite{morton2023recommendation} recommends the use of \triceratops in favor of \vespa since the latter is no longer maintained and has not been updated to account for the modern astronomy landscape. We present validation using both software packages as an independent check on one another but emphasize the results of \triceratops over those of \vespa in cases where their FPP values may disagree. This means that many of the targets in our best-in-class sample that are classified as ``Likely Planets'' may actually fall within the realm of true statistical validation when considering only the results from \triceratops. 

We also note that our statistical validation analysis cannot rule out the scenario in which validated planets with $R_{\rm p}$ $>$ 9 \rearth are actually brown dwarfs. A measured mass is required to disentangle the brown dwarf and planet scenarios and we encourage follow-up on all validated planets to this effect.

\section{Results} \label{sec:results}

\begin{figure*}
    \centering
    \includegraphics[width=0.99\textwidth]{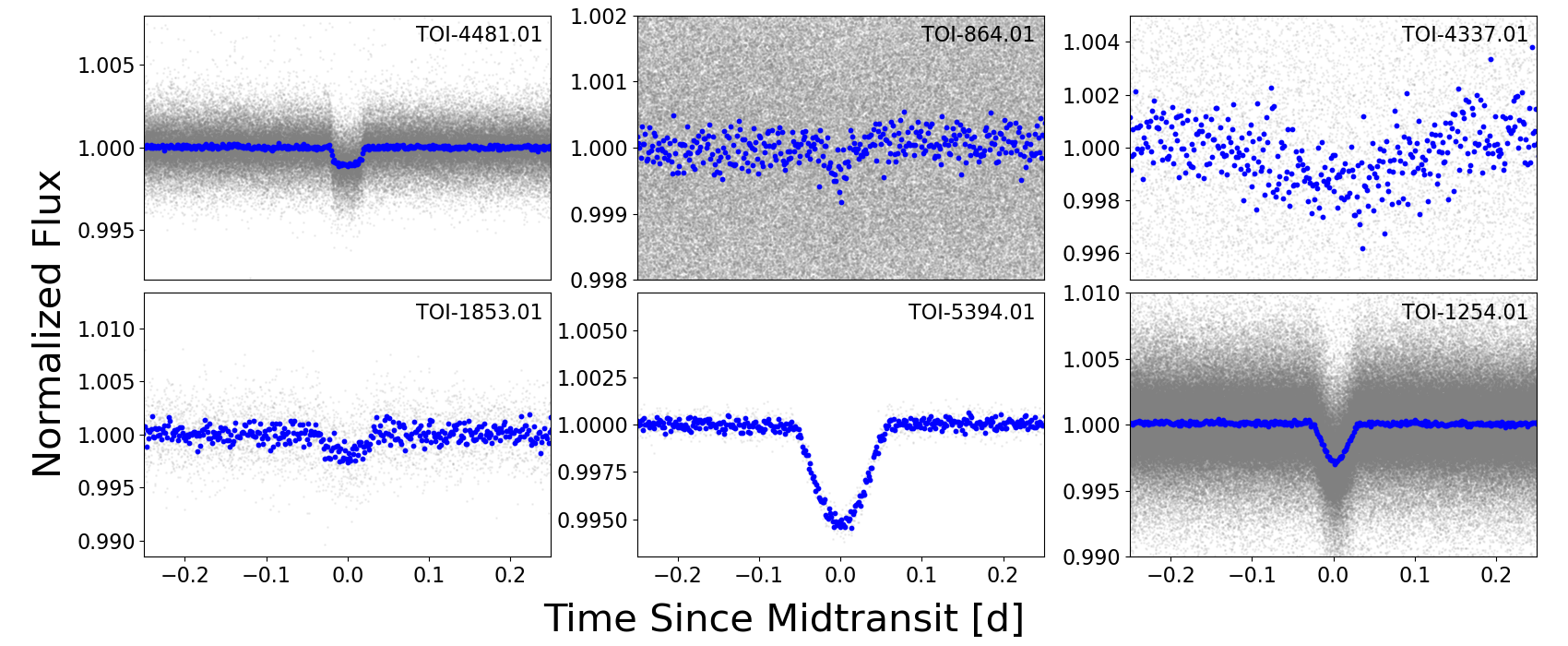}
    \caption{Examples of transits from targets in each disposition category. \textit{Left column:} examples of validated planets. Both transits are well-defined with flat bottoms. \textit{Middle column:} examples of marginal validations; a likely planet (\textit{top}) and a potential false positive (\textit{bottom}). These targets either have a low signal-to-noise ratio or a transit shape that can be confused with an eclipsing binary and cannot be validated but are also not clear false positives. \textit{Right column:} examples of likely false positives. These targets either have a very obvious V-shape, otherwise non-transit-shaped feature, or have been deemed likely false positives during vetting (e.g. a large centroid offset).}
    \label{fig:transit_examples}
\end{figure*}

Of the 103 unconfirmed \ac{TESS} planet candidates contained in our best-in-class samples, 19 passed vetting and were calculated to have \ac{FPP} values firmly meeting the threshold for statistical validation from both \vespa and \triceratops. Additionally, 11 of the original 103 unconfirmed planet candidates reside in potential multiplanet systems (TOI-1468.01, TOI-1468.02, TOI-1798.02, TOI-1806.01, TOI-2134.01, TOI-3353.01, TOI-406.01, TOI-4443.01, TOI-4495.01, TOI-836.02, TOI-880.02). Three of these have already been confirmed by independent teams (TOI-1468.01 and .02 and TOI-836.02). The remaining eight \acp{TOI} are able to take advantage of a ``multiplicity boost'' to drive their \ac{FPP} values lower. It has been shown that transit-like signals in systems with multiple transit-like signals are more likely to be true planets, assuming false positives are uniformly distributed throughout the sky \citep{lissauer2012multiplicity}. This results in a decreased \ac{FPP} value of up to 54$\times$ depending on the size of the planets, how crowded the field is for signals detected with \ac{TESS}, and the pipeline with which they were detected \citep{guerrero2021tess}. Additionally, these potential multiplanet systems represent an excellent opportunity to perform comparative planetology with the other planets in their system using \ac{JWST}.

We applied this multiplicity boost to each of the eight candidates listed above resulting in \ac{FPP} values below the validation threshold for each of them. Four of these eight already possessed \ac{FPP} values from \vespa and \triceratops that were low enough to be statistically validated, but the \ac{FPP} values of the other four targets (TOI-880.02, TOI-1798.02, TOI-1806.01, and TOI-4443.01) moved from the ``marginal validation'' range into the ``validated planet'' range. We thus arrive at a total of \vp statistically validated planets. These targets are shown in Table \ref{tab:validated_pls}. We strongly recommend these targets for additional, in-depth study and confirmation to measure their masses and model their orbits and atmospheres in preparation for potential observation with \ac{JWST}.

\begin{deluxetable}{lrrrrl}
\tablecaption{All of the statistically validated planets in both the transmission and emission spectroscopy best-in-class samples. Empty values for the TSM and ESM indicate that the target was not considered best-in-class for transmission or emission spectroscopy, respectively.}
\label{tab:validated_pls}
\tablehead{\colhead{Planet Name} & \colhead{$T_{\rm eq}$ [K]} & \colhead{$R_{\rm p}$ [\rearth]} & \colhead{Period [d]} & \colhead{TSM} & \colhead{ESM}}
\startdata
TOI-128.01 & 1345 & 2.22 & 4.94 & 90 &  \\
TOI-238.01 & 1454 & 1.86 & 1.27 & \multicolumn{1}{l}{} & \multicolumn{1}{r}{8} \\
TOI-261.01 & 1722 & 3.04 & 3.36 & 79 &  \\
TOI-332.01 & 1946 & 3.28 & 0.78 & 62 & \multicolumn{1}{r}{11} \\
TOI-406.01 & 344 & 1.96 & 13.18 & 55 &  \\
TOI-654.01 & 749 & 2.37 & 1.53 & \multicolumn{1}{l}{} & \multicolumn{1}{r}{9} \\
TOI-880.02 & 1163 & 2.78 & 2.57 & 119 & \\
TOI-907.01 & 1847 & 9.62 & 4.58 & \multicolumn{1}{l}{} & \multicolumn{1}{r}{28} \\
TOI-1135.01 & 1074 & 9.34 & 8.03 & 243 &  \\
TOI-1194.01 & 1405 & 8.89 & 2.31 & 217 & \multicolumn{1}{r}{69} \\
TOI-1347.01 & 1793 & 2.70 & 0.85 & 79 & \multicolumn{1}{r}{12} \\
TOI-1410.01 & 1396 & 2.94 & 1.22 & 118 & \multicolumn{1}{r}{20} \\
TOI-1683.01 & 929 & 2.64 & 3.06 & 101 &  \\
TOI-1798.02 & 2122 & 1.41 & 0.44 & & 6 \\
TOI-1806.01 & 337 & 3.41 & 15.15 & 60 & \\
TOI-2134.01 & 676 & 3.10 & 9.23 & 196 & \\
TOI-3353.01 & 1264 & 2.67 & 4.67 & 90 &  \\
TOI-4443.01 & 1639 & 1.72 & 1.85 & 97 & \\
TOI-4495.01 & 1383 & 3.63 & 5.18 & 74 & \\
TOI-4527.01 & 1363 & 0.91 & 0.40 & \multicolumn{1}{l}{} & \multicolumn{1}{r}{13} \\
TOI-4602.01 & 1380 & 2.55 & 3.98 & 111 & \multicolumn{1}{r}{11} \\
TOI-5082.01 & 1165 & 2.55 & 4.24 & 160 & \multicolumn{1}{r}{15} \\
TOI-5388.01 & 601 & 1.89 & 2.59 & \multicolumn{1}{l}{} & \multicolumn{1}{r}{12}
\enddata
\end{deluxetable}

\vspace{-12pt}

A total of \falsep targets were deemed ``likely false positives'' (LFPs). These targets all exhibited clear signs of a false positive in the vetting stage and/or produced \ac{FPP} values from both statistical validation software packages. A target was deemed a likely false positive if the \ac{FPP} from both \vespa and \triceratops did not meet either the validation or marginal validation thresholds. For one of these likely false positive targets, we were unable to locate the transit-like event that was flagged by the \ac{TESS} \ac{SPOC} during our manual inspection of the phase-folded light curve and we deemed it a false alarm (TOI-1022.01). Most of these \falsep likely false positive targets exhibited obvious V-shaped transits indicative of an EB and a subset of them were revealed by TFOP follow-up to have a nearby ($\leq$2'') companion star that served as the likely cause of the signal.

There was a subset of targets with high \ac{FPP} values that could be large grazing planets or systems with a high planetary to stellar radius ratio ($R_{\rm p}$/$R_{*}$) rather than their current LFP classification. Grazing transits or high $R_{\rm p}$/$R_{*}$ systems often produce transits that look somewhat V-shaped and can masquerade as a stellar eclipse rather than a planet transit. These scenarios are limiting cases for the validation software since the analyses rely so heavily on transit shape. Therefore, targets with high \ac{FPP} values that could potentially fall under these categories warrant further follow-up. For our purposes, we keep these targets classified as LFPs not only for the sake of a uniform analysis, but also because grazing transits are non-ideal candidates for transit and eclipse spectroscopy. However, we flag them here for future study and as examples of the limitations of statistical validation.

A third category of validation emerged for targets with \ac{FPP} values that did not quite meet the threshold for validation but also were not clear false positives. These \marginal targets were deemed to be marginal validations and had at least one or both \ac{FPP} values from \vespa and \triceratops that met the marginal validation criteria described in Section \ref{sec:validation}. This category was further subdivided into ``likely planets'' (LPs) and ``potential false positives'' (pFPs). LPs were targets with either both \ac{FPP} values residing in the marginal validation zone or one \ac{FPP} in the marginal validation zone and the other meeting the threshold for validation. pFPs were targets with one marginal validation \ac{FPP} and one \ac{FPP} that indicates a false positive.

The results of our vetting analysis agree with these distinctions based on \ac{FPP}. Almost all of the targets in the pFP category had at least one vetting factor that could indicate a false positive origin (e.g. V-shaped transit, possible odd-even transit depth differences, etc.) but are not definitive enough to warrant labeling the target a likely false positive. There were a total of \lp targets in the LP category and \pfp in the pFP category of marginal validations. We encourage future study and follow-up of these targets to ascertain their true nature as they could potentially be prime targets for atmospheric characterization with \ac{JWST}. Examples of transits from each disposition category are shown in Figure \ref{fig:transit_examples}.

The remaining \inconclusive targets produced inconclusive vetting and validation results. This category is distinct from marginal validations in that in most of these inconclusive cases \vespa and \triceratops disagree significantly on the status of each target or there are additional factors precluding an adequate vetting or validation analysis. The targets TOI-1355.01, TOI-1954.01, and TOI-4552.01 were validated by one statistical validation software while the other software produced an \ac{FPP} that did not meet the threshold for even a marginal validation. 

In the case of TOI-1355.01, the discrepant \acp{FPP} may be due to overly constraining photometric follow-up observations. The transit shape is slightly V-shaped, and our follow-up observations rule out a large portion of parameter EB and BEB parameter space, but those are the models that fit the phase-folded transit the best (resulting in \ac{FPP} values with large uncertainties from \triceratops). This target may be a grazing planet, which would explain the V-shape as well as the small parameter space for EBs and BEBs.

In the case of TOI-1954.01, very little follow-up exists and the target is in a crowded field, both of which likely combine to cause the discrepancy between \vespa and \triceratops. For TOI-4552.01, the signal is shallow and the light curve exhibits some variability which is likely causing variability in the \ac{FPP} values calculated by the different validation software packages.

The final inconclusive case is TOI-4597.01. This target was statistically validated by \vespa but \triceratops was unable to run on it. This is likely due to the short periodic oscillations that appear in the light curve as a result of stellar activity or variability. A clear transit exists, but we are unable to complete our vetting and validation analysis without properly modeling the variability in the light curve to produce a clean transit. This is beyond the scope of this work as it would require a physically-motivated model to subtract from the light curve that our vetting and validation procedure is incapable of. We encourage follow-up analysis of these four inconclusive targets to determine their true nature.

\begin{figure*}
    \centering
    \includegraphics[width=0.95\textwidth]{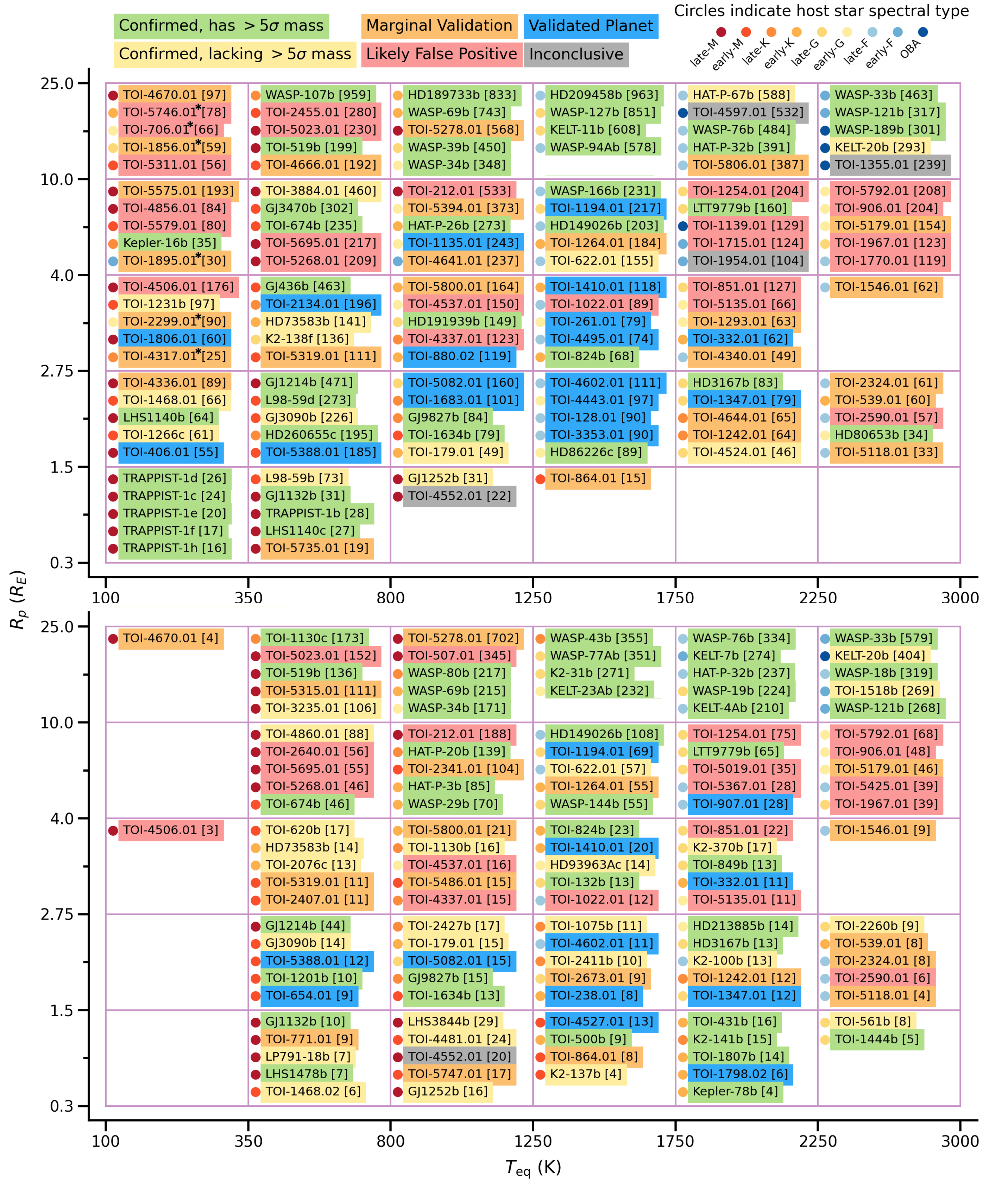}
    \caption{Our best-in-class targets for transmission (\textit{top}) and emission (\textit{bottom}) spectroscopy after performing our vetting and validation analysis on the sample. Similar to Figure \ref{fig:original_grids}, target names are displayed in the cell corresponding to the parameter space they occupy next to their 
    \ac{TSM} or \ac{ESM} value in brackets with approximate host stellar type denoted by the colored circle, as determined by reported effective temperature. Each target's background color corresponds to its mass measurement and validation status: green targets are confirmed planets with mass measurements $>$5$\sigma$, yellow targets are confirmed planets with mass measurements $<$5$\sigma$ and \acp{TOI} that were independently confirmed over the course of our analysis, blue targets have been statistically validated by our analysis, orange targets are marginal validations (LPs and pFPs), red targets were deemed likely false positives (LFPs) by our analysis, and gray targets were deemed to have an inconclusive validation. Targets with an asterisk next to their name have an ambiguous period and their \ac{TSM} values are liable to change as their $T_{\rm{eq}}$ values were calculated on the reported periods not the true periods.}
    \label{fig:new_grids}
\end{figure*}

There were an additional \confirmed targets from our samples that were confirmed by independent teams over the course of our analysis. These targets are TOI-179~b \citep{desidera2022toi179,deleon2023hd18599}, TOI-622~b \citep{psaridi2023toi615toi622toi2641}, TOI-836~b \citep{hawthorn2023toi836}, TOI-969~b \citep{lillobox2023toi969}, TOI-1099~b \citep{barros2023toi1099}, TOI-1468~b~$\&$~c \citep{chaturvedi2022toi1468}, TOI-1853~b \citep{naponielloaccepted}, TOI-3235~b \citep{hobson2023toi3235}, TOI-3884~b \citep{almenara2022toi3884}, TOI-4463~A~b \citep{yee2023unifiedhjsurvey2}, GJ~806~b \citep[TOI-4481~b;][]{palle2023gj806}, HD~20329~b \citep[TOI-4524~b][]{murgas2022hd20329}, and TOI-4860~b \citep{almenara2023toi4860,triaud2023toi4860}. This high number of targets in our best-in-class sample being confirmed in such a short period of time is very positive for the prospects for atmospheric characterization with JWST. Indeed, the goal of our synthesis of the best-in-class sample is to highlight and elevate targets potentially well-suited to such observations for follow-up to measure their masses and confirm their planetary nature.

For the targets that had masses measured independently over the course of our analysis, we recalculated their \ac{TSM} and \ac{ESM} values according to their updated planet parameters and reranked them within their respective bins. In the cases of TOI-179 b, TOI-836 b, TOI-969 b, TOI-1099 b, TOI-1853 b, TOI-3235 b, TOI-4463 A b, and TOI-4860 b, the updated parameters differed significantly from those originally listed on the NASA Exoplanet Archive, resulting in \ac{TSM} and \ac{ESM} values much lower than calculated and warranting removal from either the best-in-class transmission or emission spectroscopy samples. In the cases of TOI-836 b, TOI-969 b, TOI-1099 b, TOI-1853 b, and TOI-4463 A b, this amounted to removal from the entire best-in-class sample.

Our best-in-class sample also includes six targets with ambiguous periods -- TOI-706.01, TOI-1856.01, TOI-1895.01, TOI-2299.01, TOI-4317.01, TOI-5575.01, and TOI-5746.01. These targets were originally discovered as single transits before transiting again in later sectors of \ac{TESS} data. The periods reported on ExoFOP represent the upper limit on their periods since additional transits of these targets could have fallen in gaps in the \ac{TESS} data and their true periods may be shorter. We performed our vetting and validation using the stated periods but knowing that future observations could reveal shorter periods that would alter the \ac{TSM} and \ac{ESM} values as well as their observability with \ac{JWST}. We choose not to discard these targets from our best-in-class samples to emphasize their potential as ideal \ac{JWST} targets and emphasize the need for additional follow-up on them. Only TOI-4317.01 had low enough \ac{FPP} values to be considered statistically validated, but due to its ambiguous period, we place it in the ``likely planet'' category and we caution that a deeper analysis is required for this target to identify the true period and therefore its true planetary status. We note that the planets with long or ambiguous periods in our sample should have their orbital periods further scrutinized as the \ac{TESS} observing strategy makes it difficult to determine such long orbital periods and they may change depending on individual circumstances.

The final best-in-class sample is displayed in Figure \ref{fig:new_grids}, which mirrors Figure \ref{fig:original_grids} but now includes updated dispositions for all targets in our samples, both confirmed and unconfirmed. Additional information on each target can be found in Table \ref{tab:full_sample} in Appendix \ref{app:full_table}. An extended machine-readable version of this table is also available in the online version of this article.

There were a number of targets in our best-in-class sample that are also in the process of being validated by independent teams. These include TOI-4226.01 \citep{timmermanstoi4336inprep} and TOI-4317.01 \citep{osborninprep}. We direct the reader to these upcoming publications for a more in-depth analysis and exploration than is available here and to treat such in-depth analyses as the definitive discovery papers for these individual targets. \cite{dressinginprep} is also conducting a parallel large-scale validation effort on TOIs 261.01, 4317.01, 4527.01, 4602.01, and 5082.01, as is \cite{mistryinprep} for TOIs 238.01 and 771.01 and we direct the reader to this upcoming publication for an additional, independent vetting and validation of these targets. Additionally, independent teams are conducting confirmation and characterization of TOI-1410.01 \citep{livingstoninprep}, TOIs 1194.01, 1347.01, and 1410.01 \citep{polanskiinprep}, and TOI-880.02 \citep{nielseninprep} and we direct the reader to these papers for an in-depth analysis of these targets.

\section{Summary and Conclusion} \label{sec:summary}

In this paper, we present a set of best-in-class planets for atmospheric characterization with \ac{JWST} through both transmission and emission spectroscopy. Our vetting, validation, and results are summarized here:

\begin{itemize}
    \item We queried the NASA Exoplanet Archive for all transiting confirmed planets and unconfirmed \ac{TESS} candidates and calculated their TSM, ESM, and observability with \ac{JWST}.
    \item We divided all planets into grids with bins in equilibrium temperature, $T_{\rm eq}$, from 100 to 3000 K and planetary radius, $R_{\rm p}$, from 0.3 to 25.0 \rearth and the top five planets and candidates were ranked by spectroscopy metric in each bin to create a best-in-class sample for each spectroscopy method.
    \item The 103 unconfirmed \ac{TESS}-discovered candidates from the transmission and emission spectroscopy grids were vetted using a combination of follow-up observations collected by TFOP and independent analyses such as the \ac{SPOC} DV reports.
    \item We used \vespa and \triceratops to calculate the false positive probabilities and determine a final disposition for each target.
    \item Our analysis resulted in \vp validated targets, \falsep likely false positives, \marginal targets that were marginally validated, and \inconclusive inconclusive validations. Of our original targets, \confirmed were independently confirmed over the course of our analysis.
    \item This final sample represents the best-in-class targets for atmospheric characterization with \ac{JWST} and deeper analysis on each target is highly encouraged.
\end{itemize}

The best-in-class sample presented in this paper is meant to represent an initial look at many of the targets with the potential to yield high quality spectra from \ac{JWST}. We hope that this work paves the way for future studies of a similar sort. We highly encourage independent analysis of each target presented here to discern the true nature of each and build a catalog of planets that can reliably provide exquisite atmospheric data from \ac{JWST}.

We recognize that this best-in-class sample will undoubtedly change over time as new targets supplant previous ones in the \ac{JWST} observability rankings, as targets are shown to be false positives, or as the orbital and planetary parameters of targets are refined with further observation. Since the date on which we queried the NASA Exoplanet Archive to generate our sample, $\sim$750 new \ac{TESS} candidates have been discovered. It is possible that anywhere from 5-10 of these new discoveries could belong possess \ac{TSM} or \ac{ESM} values that place them within the best-in-class sample. These targets primarily appear in the bins containing the largest planets and the hottest planets. As \ac{TESS} probes fainter stars, new detections are even more biased towards large, hot planets as they possess a sufficient signal-to-noise to be detected around faint stars.

This sample may also change based on the assumptions used to generate it. Our analysis calculated the \ac{ESM} value for planets in all portions of parameter space even though it was originally developed by \cite{kempton2018tsmesm} for terrestrial planets. Parameter values baked into the \ac{ESM} quantity such as the day-night heat redistribution on a planet may be different from what is assumed by our calculations. However, since our rankings of best-in-class targets are relative to other planets and candidates of similar radius and equilibrium temperature, this factor can likely be ignored. Furthermore, the discrete boundaries of our bins may bias our best-in-class sample towards targets at the hot and large edges of their bins, so different binning schemes may change the specific targets that are contained within the best-in-class sample.

Additionally, it is possible that the thermal emission of planets hotter than $\sim$800~K can be observed with NIR instruments rather than with MIRI as assumed by our analysis. This would open up access to brighter stars due to the favorable ratio between the flux of the planet's thermal emission and the flux of the star and would allow for study of a different set of spectral features compared to those available to MIRI. The sample presented here makes parameter cuts for emission spectroscopy based on the performance of MIRI, but a blend of instruments would open up the pool of potential best-in-class targets for the hottest portions of parameter space.

This best-in-class sample may also provide useful for future missions that will study exoplanet atmospheres, such as the upcoming Ariel mission which will conduct a survey of around one thousand exoplanetary atmospheres. A total of 69 of our best-in-class targets are contained within the Ariel target list described by \cite{edwards2022ariel}. This overlap may grow as both our best-in-class sample and the Ariel target list are updated.

To a first order approximation, out of 103 total targets originally unconfirmed in our best-in-class sample, 52 of them were either statistically validated, marginally validated and ruled ``likely planets'', or were confirmed independently. This suggests that at least $\sim$50$\%$ of the \ac{TESS} candidates analyzed are true planets, although this value may be higher if any of the targets deemed ``potential false positive'' or ``likely false positive'' are actually planets.

This sample also demonstrates the power of \ac{TESS} to discover planets amenable for atmospheric characterization from which we can learn a great deal about their atmospheric structure and composition. Approximately 57$\%$ of the targets in the final best-in-class sample (excluding likely false positives) are \ac{TESS} discoveries. However, \ac{TESS} has surprisingly missed the detection of some small planets orbiting small stars, so planet searches beyond \ac{TESS} are also required \citep{brady2022tessmdwarfplanets}. It is therefore important to continue searching for planet candidates that could turn out to be excellent targets for atmospheric study since, as shown here, many of the best planets for study with \ac{JWST} are still being revealed.

\vspace{24pt}

We thank T. Komacek, D. Richardson, A. Boss, and C. Hartzell for their helpful discussion of this work. We also thank E. Petigura for his contributions to HIRES observations supporting this work.

Funding for the TESS mission is provided by NASA's Science Mission Directorate. 

This research has made use of the Exoplanet Follow-up Observation Program (ExoFOP; DOI: 10.26134/ExoFOP5) website, which is operated by the California Institute of Technology, under contract with the National Aeronautics and Space Administration under the Exoplanet Exploration Program.

This work has made use of data from the European Space Agency (ESA) mission {\it Gaia} (\url{https://www.cosmos.esa.int/gaia}), processed by the {\it Gaia} Data Processing and Analysis Consortium (DPAC, \url{https://www.cosmos.esa.int/web/gaia/dpac/consortium}). Funding for the DPAC has been provided by national institutions, in particular the institutions participating in the {\it Gaia} Multilateral Agreement.

This paper includes data collected by the TESS mission, which are publicly available from the Mikulski Archive for Space Telescopes (MAST) and produced by the Science Processing Operations Center (SPOC) at NASA Ames Research Center. This research effort made use of systematic error-corrected (PDC-SAP) photometry. Funding for the TESS mission is provided by NASA's Science Mission directorate. 

Resources supporting this work were provided by the NASA High-End Computing (HEC) Program through the NASA Advanced Supercomputing (NAS) Division at Ames Research Center for the production of the SPOC data products.

This work makes use of observations from the LCOGT network. Part of the LCOGT telescope time was granted by NOIRLab through the Mid-Scale Innovations Program (MSIP). MSIP is funded by NSF.

This paper is based on observations made with the MuSCAT3 instrument, developed by the Astrobiology Center and under financial supports by JSPS KAKENHI (JP18H05439) and JST PRESTO (JPMJPR1775), at Faulkes Telescope North on Maui, HI, operated by the Las Cumbres Observatory.

This paper makes use of observations made with the MuSCAT2 instrument, developed by the Astrobiology Center, at TCS operated on the island of Tenerife by the IAC in the Spanish Observatorio del Teide.

This paper makes use of data from the MEarth Project, which is a collaboration between Harvard University and the Smithsonian Astrophysical Observatory. The MEarth Project acknowledges funding from the David and Lucile Packard Fellowship for Science and Engineering, the National Science Foundation under grants AST-0807690, AST-1109468, AST-1616624 and AST-1004488 (Alan T. Waterman Award), the National Aeronautics and Space Administration under Grant No. 80NSSC18K0476 issued through the XRP Program, and the John Templeton Foundation.

C. M. would like to gratefully acknowledge the entire Dragonfly Telephoto Array team, and Bob Abraham in particular, for allowing their telescope bright time to be put to use observing exoplanets.

B.J.H. acknowledges support from the Future Investigators in NASA Earth and Space Science and Technology (FINESST) program - grant 80NSSC20K1551 - and support by NASA under award number 80GSFC21M0002.

K.A.C. and C.N.W. acknowledge support from the TESS mission via subaward s3449 from MIT.

This research made use of Lightkurve, a Python package for Kepler and TESS data analysis \citep{lightkurve2018lightkurve}.

Some of the data presented herein were obtained at the W. M. Keck Observatory, which is operated as a scientific partnership among the California Institute of Technology, the University of California and the National Aeronautics and Space Administration. The Observatory was made possible by the generous financial support of the W. M. Keck Foundation. The authors wish to recognize and acknowledge the very significant cultural role and reverence that the summit of Maunakea has always had within the indigenous Hawaiian community.  We are most fortunate to have the opportunity to conduct observations from this mountain.

D.R.C. and C.A.C. acknowledge support from NASA through the XRP grant 18-2XRP18$\_$2-0007. C.A.C. acknowledges that this research was carried out at the Jet Propulsion Laboratory, California Institute of Technology, under a contract with the National Aeronautics and Space Administration (80NM0018D0004).

This research was carried out in part at the Jet Propulsion Laboratory, California Institute of Technology, under a contract with the National Aeronautics and Space Administration (80NM0018D0004).

S.Z. and A.B. acknowledge support from the Israel Ministry of Science and Technology (grant No. 3-18143).

The research leading to these results has received funding from  the ARC grant for Concerted Research Actions, financed by the Wallonia-Brussels Federation. TRAPPIST is funded by the Belgian Fund for Scientific Research (Fond National de la Recherche Scientifique, FNRS) under the grant PDR T.0120.21. TRAPPIST-North is a project funded by the University of Liege (Belgium), in collaboration with Cadi Ayyad University of Marrakech (Morocco). M.G. is F.R.S.-FNRS Research Director and E.J. is F.R.S.-FNRS Senior Research Associate. The postdoctoral fellowship of K.B. is funded by F.R.S.-FNRS grant T.0109.20 and by the Francqui Foundation.

H.P.O.’s contribution has been carried out within the framework of the NCCR PlanetS supported by the Swiss National Science Foundation under grants 51NF40\_182901 and 51NF40\_205606.

F.J.P. acknowledges financial support from the grant CEX2021-001131-S funded by MCIN/AEI/ 10.13039/501100011033.

A.J. acknowledges support from ANID -- Millennium Science Initiative -- ICN12\_009 and from FONDECYT project 1210718.

Z.L.D acknowledges the MIT Presidential Fellowship and that this material is based upon work supported by the National Science Foundation Graduate Research Fellowship under Grant No. 1745302.

Some of the observations in this paper made use of the High-Resolution Imaging instruments ‘Alopeke and Zorro, and were obtained under Gemini LLP Proposal Number: GN/S-2021A-LP-105. ‘Alopeke/Zorro were funded by the NASA Exoplanet Exploration Program and built at the NASA Ames Research Center by Steve B. Howell, Nic Scott, Elliott P. Horch, and Emmett Quigley. Alopeke/Zorro was mounted on the Gemini North/South 8-m telescopes of the international Gemini Observatory, a program of NSF’s OIR Lab, which is managed by the Association of Universities for Research in Astronomy (AURA) under a cooperative agreement with the National Science Foundation. on behalf of the Gemini partnership: the National Science Foundation (United States), National Research Council (Canada), Agencia Nacional de Investigación y Desarrollo (Chile), Ministerio de Ciencia, Tecnología e Innovación (Argentina), Ministério da Ciência, Tecnologia, Inovações e Comunicações (Brazil), and Korea Astronomy and Space Science Institute (Republic of Korea).

This work is partly supported by JSPS KAKENHI Grant Numbers JP17H04574, JP18H05439, JP21K20376, JST CREST Grant Number JPMJCR1761, and Astrobiology Center SATELLITE Research project AB022006.

This article is based on observations made with the MuSCAT2 instrument, developed by ABC, at Telescopio Carlos Sánchez operated on the island of Tenerife by the IAC in the Spanish Observatorio del Teide. This paper is based on observations made with the MuSCAT3 instrument, developed by the Astrobiology Center and under financial supports by JSPS KAKENHI (JP18H05439) and JST PRESTO (JPMJPR1775), at Faulkes Telescope North on Maui, HI, operated by the Las Cumbres Observatory.

This publication benefits from the support of the French Community of Belgium in the context of the FRIA Doctoral Grant awarded to M.T.

D. D. acknowledges support from TESS Guest Investigator Program grants
80NSSC22K1353, 80NSSC22K0185 and 80NSSC23K0769.

A.B. acknowledges the support of M.V. Lomonosov Moscow State University Program of Development.

T.D. was supported in part by the McDonnell Center for the Space Sciences.

\facilities{Adams Observatory, ASP, Brierfield Private Observatory, Campo Catino Astronomical Observatory, Catania Astrophysical Observatory, Caucasian Mountain Observatory, CHAT, CROW Observatory, Deep Sky West, Dragonfly, El Sauce, ExTrA, FEROS, Fred L. Whipple Observatory, Gaia, Gemini (‘Alopeke, Zorro), George Mason University, HATSouth, Hazelwood Observatory, Keck, Kutztown University Observatory, LCOGT, Lewin Observatory, Lick Observatory, Lookout Observatory, MASTER-Ural, MEarth-S, Mt. Stuart Observatory, MuSCAT, MuSCAT2, MuSCAT3, Observatori Astron\`{o}mic de la Universitat de Val\`{e}ncia, Observatori Astron\`{o}mic Albany\`{a}, Observatorio del Roque de los Muchachos, Observatory de Ca l'Ou, Palomar Observatory, PEST, Privat Observatory Herges-Hallenberg, PvDKO, RCO, SAINT-EX, Salerno University Observatory, Solaris SLR2, SPECULOOS, SUTO-Otivar, TESS, TRAPPIST, University of Louisville telescopes at the University of Southern Queensland's Mt. Kent Observatory and at Mt. Lemmon Observatory, Union College Observatory, Villa '39, VLT, WASP, WCO, Wellesley College Whitin Observatory, WIYN}

\software{\texttt{AstroImageJ}, Astropy \citep{astropy:2013, astropy:2018, astropy:2022}, astroquery \citep{astroquery:2019}, \texttt{BANZAI} \citep{McCully:2018}, \texttt{DEATHSTAR}, Jupyter \citep{kluyver2016jupyter}, Lightkurve \citep{lightkurve2018lightkurve}, matplotlib \citep{hunter2007matplotlib}, numpy \citep{van2011numpy}, pandas \citep{pandas}, \texttt{TESS Transit Finder} \citep{Jensen:2013}, Tesscut \citep{Tesscut:2019}, TRICERATOPS \citep{giacalone2020triceratops, giacalone2021triceratopsvetting}, VESPA \citep{morton2012vespa, morton2015vespa}}

\bibliography{main}{}
\bibliographystyle{aasjournal}

\appendix

\section{Full List of Vetted and Validated Best-in-Class TOIs} \label{app:full_table}

\input{appendix_table}

\section{List of TFOP Observations Used in Vetting and Validation of the Best-in-Class Sample} \label{app:tfop_obs}

\input{tfop_obs_table_short}

\end{document}

%% file: authors.tex
\author[0000-0001-5084-4269]{Benjamin J. Hord}
\affiliation{Department of Astronomy, University of Maryland, College Park, MD 20742, USA}
\affiliation{NASA Goddard Space Flight Center, 8800 Greenbelt Road, Greenbelt, MD 20771, USA}
\affiliation{GSFC Sellers Exoplanet Environments Collaboration}

\author[0000-0002-1337-9051]{Eliza M.-R. Kempton}
\affiliation{Department of Astronomy, University of Maryland, College Park, MD 20742, USA}

\author[0000-0001-5442-1300]{Thomas Mikal-Evans}
\affiliation{Max Planck Institute for Astronomy, K\"{o}nigstuhl 17, D-69117 Heidelberg, Germany}

\author[0000-0001-9911-7388]{David W. Latham}
\affiliation{Center for Astrophysics \textbar \ Harvard \& Smithsonian, 60 Garden Street, Cambridge, MA 02138, USA}

\author[0000-0002-5741-3047]{David R. Ciardi}
\affiliation{NASA Exoplanet Science Institute, IPAC, California Institute of Technology, Pasadena, CA 91125 USA}

\author[0000-0003-2313-467X]{Diana Dragomir}
\affiliation{Department of Physics and Astronomy, University of New Mexico, 210 Yale Boulevard NE, Albuquerque, NM 87106, USA}

\author[0000-0001-8020-7121]{Knicole D. Col\'{o}n}
\affiliation{NASA Goddard Space Flight Center, 8800 Greenbelt Road, Greenbelt, MD 20771, USA}
\affiliation{GSFC Sellers Exoplanet Environments Collaboration}

\author{Gabrielle Ross}
\affiliation{Department of Earth, Atmospheric and Planetary Sciences, Massachusetts Institute of Technology, Cambridge, MA 02139, USA}

\author[0000-0001-7246-5438]{Andrew Vanderburg}
\affiliation{Department of Earth, Atmospheric and Planetary Sciences, Massachusetts Institute of Technology, Cambridge, MA 02139, USA}

\author[0000-0002-7564-6047]{Zoe\ L. de Beurs}
\altaffiliation{NSF Graduate Research Fellow and MIT Presidential Fellow}
\affiliation{Department of Earth, Atmospheric and Planetary Sciences, Massachusetts Institute of Technology, Cambridge, MA 02139, USA}

% ...


%% file: appendix_table.tex
\startlongtable
\begin{deluxetable}{cccccccccc}
\tabletypesize{\footnotesize}
\tablecaption{Our full best-in-class sample including both the \acp{TOI} on which we performed vetting and validation analysis and planets that were confirmed prior to our study or independent of our analysis. In the disposition column, VP=validated planet, LP=likely planet, pFP=possible false positive, LFP=likely false positive. See Section \ref{sec:results} for a further explanation of each disposition and their definitions. For an extended, machine-readable version of this table with additional parameters, please see the online version of this article.}
\label{tab:full_sample}
\tablehead{ \colhead{Planet Name} & \colhead{Period} & \colhead{Radius} & \colhead{$T_{\rm eq}$} & \colhead{Semi-major Axis} & \colhead{Host Star $T_{\rm eff}$} & \colhead{Host Star Mass} & \colhead{TSM} & \colhead{ESM} & \colhead{Disposition} \\
\colhead{} & \colhead{(days)} & \colhead{($R_{\rm \oplus}$)} & \colhead{(K)} & \colhead{(AU)} & \colhead{(K)} & \colhead{($M_{\rm \odot}$)} & \colhead{} & \colhead{} & \colhead{} }

\startdata
GJ 436 b & 2.64 & 3.96 & 684 & 0.029 & 3586 & 0.47 & \textbf{463} & 108 & Confirmed \\
GJ 1132 b & 1.63 & 1.13 & 578 & 0.015 & 3270 & 0.18 & \textbf{31} & \textbf{10} & Confirmed \\
GJ 1214 b & 1.58 & 2.74 & 582 & 0.015 & 3170 & 0.18 & \textbf{471} & \textbf{44} & Confirmed \\
GJ 1252 b & 0.52 & 1.19 & 1048 & 0.009 & 3325 & 0.38 & \textbf{31} & \textbf{16} & Confirmed \\
GJ 3090 b & 2.85 & 2.13 & 721 & 0.032 & 3703 & 0.52 & \textbf{226} & \textbf{14} & Confirmed \\
GJ 3470 b & 3.34 & 4.36 & 691 & 0.036 & 3652 & 0.54 & \textbf{302} & 38 & Confirmed \\
GJ 9827 b & 1.21 & 1.58 & 1184 & 0.019 & 4340 & 0.61 & \textbf{84} & \textbf{15} & Confirmed \\
HAT-P-3 b & 2.90 & 9.98 & 1157 & 0.039 & 5185 & 1.06 & 96 & \textbf{85} & Confirmed \\
HAT-P-20 b & 2.88 & 9.72 & 972 & 0.036 & 4595 & 0.76 & 13 & \textbf{139} & Confirmed \\
HAT-P-26 b & 4.23 & 7.06 & 993 & 0.048 & 5079 & 1.12 & \textbf{273} & 27 & Confirmed \\
HAT-P-32 b & 2.15 & 20.05 & 1918 & 0.034 & 6269 & 1.13 & \textbf{391} & \textbf{237} & Confirmed \\
HAT-P-67 b & 4.81 & 23.37 & 1899 & 0.065 & 6406 & 1.64 & \textbf{588} & 115 & Confirmed \\
HD 3167 b & 0.96 & 1.63 & 1772 & 0.018 & 5286 & 0.85 & \textbf{83} & \textbf{13} & Confirmed \\
HD 73583 b & 6.40 & 2.79 & 743 & 0.060 & 4695 & 0.73 & \textbf{141} & \textbf{14} & Confirmed \\
HD 80653 b & 0.72 & 1.61 & 2446 & 0.017 & 5910 & 1.18 & \textbf{34} & 6 & Confirmed \\
HD 86226 c & 3.98 & 2.16 & 1304 & 0.049 & 5863 & 1.02 & \textbf{89} & 11 & Confirmed \\
HD 93963 A c & 3.65 & 3.23 & 1344 & 0.048 & 5987 & 1.11 & 67 & \textbf{14} & Confirmed \\
HD 149026 b & 2.88 & 8.30 & 1676 & 0.043 & 6179 & 1.42 & \textbf{203} & \textbf{108} & Confirmed \\
HD 189733 b & 2.22 & 12.67 & 1201 & 0.031 & 5052 & 0.79 & \textbf{833} & 1140 & Confirmed \\
HD 191939 b & 8.88 & 3.39 & 908 & 0.078 & 5427 & 0.81 & \textbf{149} & 13 & Confirmed \\
HD 209458 b & 3.52 & 15.47 & 1451 & 0.049 & 6091 & 1.23 & \textbf{963} & 546 & Confirmed \\
HD 213885 b & 1.01 & 1.75 & 2068 & 0.020 & 5795 & 1.07 & 56 & \textbf{14} & Confirmed \\
HD 260655 c & 5.71 & 1.53 & 556 & 0.047 & 3803 & 0.44 & \textbf{195} & 9 & Confirmed \\
K2-31 b & 1.26 & 11.88 & 1556 & 0.022 & 5412 & 0.91 & 120 & \textbf{271} & Confirmed \\
K2-100 b & 1.67 & 1.52 & 1911 & 0.030 & 6168 & 1.15 & 3 & \textbf{13} & Confirmed \\
K2-137 b & 0.18 & 0.89 & 1704 & 0.006 & 3492 & 0.46 & 19 & \textbf{4} & Confirmed \\
K2-138 f & 12.76 & 2.90 & 741 & 0.104 & 5356 & 0.94 & \textbf{136} & 2 & Confirmed \\
K2-141 b & 0.28 & 1.49 & 2042 & 0.007 & 4373 & 0.71 & 8 & \textbf{15} & Confirmed \\
K2-370 b & 2.14 & 3.21 & 1986 & 0.017 & 5372 & 0.98 & 100 & \textbf{17} & Confirmed \\
KELT-4 A b & 2.99 & 19.04 & 1823 & 0.043 & 6206 & 1.20 & 310 & \textbf{210} & Confirmed \\
KELT-7 b & 2.73 & 17.18 & 2041 & 0.044 & 6768 & 1.76 & 253 & \textbf{274} & Confirmed \\
KELT-11 b & 4.74 & 13.69 & 1703 & 0.063 & 5375 & 1.44 & \textbf{608} & 139 & Confirmed \\
KELT-20 b & 3.47 & 20.58 & 2331 & 0.054 & 8980 & 1.76 & \textbf{293} & \textbf{404} & Confirmed \\
KELT-23 A b & 2.26 & 14.83 & 1562 & 0.033 & 5899 & 0.94 & 283 & \textbf{232} & Confirmed \\
Kepler-16 b & 228.78 & 8.45 & 206 & 0.705 & 4450 & 0.69 & \textbf{35} & 0 & Confirmed \\
Kepler-78 b & 0.36 & 1.23 & 2220 & 0.009 & 5121 & 0.84 & 7 & \textbf{4} & Confirmed \\
L 98-59 b & 2.25 & 0.95 & 623 & 0.022 & 3412 & 0.27 & \textbf{73} & 5 & Confirmed \\
L 98-59 d & 7.45 & 1.52 & 416 & 0.049 & 3412 & 0.27 & \textbf{273} & 4 & Confirmed \\
LHS 1140 b & 24.74 & 1.73 & 232 & 0.096 & 3216 & 0.19 & \textbf{64} & 0 & Confirmed \\
LHS 1140 c & 3.78 & 1.28 & 433 & 0.027 & 3216 & 0.19 & \textbf{27} & 3 & Confirmed \\
LHS 1478 b & 1.95 & 1.24 & 595 & 0.018 & 3381 & 0.24 & 18 & \textbf{7} & Confirmed \\
LHS 3844 b & 0.46 & 1.30 & 807 & 0.006 & 3043 & 0.15 & 41 & \textbf{29} & Confirmed \\
LP 791-18 b & 0.95 & 1.12 & 631 & 0.010 & 2960 & 0.14 & 19 & \textbf{7} & Confirmed \\
LTT 9779 b & 0.79 & 4.51 & 1955 & 0.017 & 5443 & 0.77 & \textbf{160} & \textbf{65} & Confirmed \\
TOI-128.01 & 4.94 & 2.22 & 1345 & 0.054 & 6086 & 0.85 & \textbf{90} & 8 & VP \\
TOI-132 b & 2.11 & 3.42 & 1513 & 0.026 & 5397 & 0.97 & 40 & \textbf{13} & Confirmed \\
TOI-179.01 & 4.14 & 2.60 & 969 & 0.048 & 5145 & 0.86 & 49 & \textbf{15} & Confirmed \\
TOI-212.01 & 0.34 & 4.47 & 1111 & 0.006 & 3332 & 0.29 & \textbf{533} & \textbf{188} & LFP \\
TOI-238.01 & 1.27 & 1.86 & 1454 & 0.021 & 5024 & 0.79 & 70 & \textbf{8} & VP \\
TOI-261.01 & 3.36 & 3.04 & 1722 & 0.035 & 5890 & 0.50 & \textbf{79} & 10 & VP \\
TOI-332.01 & 0.78 & 3.28 & 1946 & 0.016 & 5190 & 0.88 & \textbf{62} & \textbf{11} & VP \\
TOI-406.01 & 13.18 & 1.96 & 344 & 0.086 & 3349 & 0.48 & \textbf{55} & 1 & VP \\
TOI-431 b & 0.49 & 1.28 & 1879 & 0.011 & 4850 & 0.78 & 16 & \textbf{16} & Confirmed \\
TOI-500 b & 0.55 & 1.17 & 1683 & 0.012 & 4621 & 0.74 & 16 & \textbf{9} & Confirmed \\
TOI-507.01 & 0.90 & 17.12 & 875 & 0.017 & 3338 & 0.82 & 291 & \textbf{345} & LFP \\
TOI-519 b & 1.27 & 11.5 & 751 & 0.016 & 3322 & 0.34 & \textbf{199} & \textbf{136} & Confirmed \\
TOI-539.01 & 0.31 & 1.52 & 2370 & 0.008 & 4836 & 0.68 & \textbf{60} & \textbf{8} & LP \\
TOI-561 b & 0.45 & 1.42 & 2371 & 0.011 & 5455 & 0.79 & 19 & \textbf{8} & Confirmed \\
TOI-620 b & 5.10 & 3.76 & 604 & 0.048 & 3708 & 0.58 & 171 & \textbf{17} & Confirmed \\
TOI-622.01 & 6.40 & 9.24 & 1388 & 0.071 & 6400 & 1.31 & \textbf{155} & \textbf{57} & Confirmed \\
TOI-654.01 & 1.53 & 2.37 & 749 & 0.021 & 3433 & 0.53 & 78 & \textbf{9} & VP \\
TOI-674 b & 1.98 & 5.25 & 695 & 0.025 & 3514 & 0.42 & \textbf{235} & \textbf{46} & Confirmed \\
TOI-706.01 & 719.04 & 16.63 & 333 & 1.024 & 5710 & 0.28 & \textbf{66} & 3 & LFP \\
TOI-771.01 & 2.33 & 1.40 & 663 & 0.013 & 3231 & 0.06 & 18 & \textbf{9} & pFP \\
TOI-824 b & 1.39 & 2.93 & 1254 & 0.022 & 4600 & 0.71 & \textbf{68} & \textbf{23} & Confirmed \\
TOI-849 b & 0.77 & 3.44 & 1975 & 0.016 & 5374 & 0.93 & 21 & \textbf{13} & Confirmed \\
TOI-851.01 & 0.63 & 3.39 & 1954 & 0.014 & 5485 & 0.94 & \textbf{127} & \textbf{22} & LFP \\
TOI-864.01 & 0.52 & 1.00 & 1272 & 0.007 & 3460 & 0.16 & \textbf{15} & \textbf{8} & LP \\
TOI-880.02 & 2.57 & 2.78 & 1163 & 0.034 & 4935 & 0.81 & \textbf{119} & 15 & VP \\
TOI-906.01 & 1.66 & 5.05 & 2450 & 0.019 & 5955 & 0.36 & \textbf{204} & \textbf{48} & LFP \\
TOI-907.01 & 4.58 & 9.62 & 1847 & 0.055 & 6272 & 1.07 & 91 & \textbf{28} & VP \\
TOI-1022.01 & 3.10 & 3.43 & 1715 & 0.041 & 6084 & 0.98 & \textbf{89} & \textbf{12} & LFP \\
TOI-1075 b & 0.60 & 1.72 & 1325 & 0.012 & 3921 & 0.60 & 72 & \textbf{11} & Confirmed \\
TOI-1130 b & 4.07 & 3.65 & 810 & 0.044 & 4250 & 0.68 & 127 & \textbf{16} & Confirmed \\
TOI-1130 c & 8.35 & 11.14 & 638 & 0.071 & 4250 & 0.68 & 106 & \textbf{173} & Confirmed \\
TOI-1135.01 & 8.03 & 9.34 & 1074 & 0.082 & 5963 & 1.16 & \textbf{243} & 51 & VP \\
TOI-1139.01 & 4.48 & 7.61 & 2013 & 0.066 & 7947 & 1.93 & \textbf{129} & 24 & LFP \\
TOI-1194.01 & 2.31 & 8.89 & 1405 & 0.033 & 5340 & 0.92 & \textbf{217} & \textbf{69} & VP \\
TOI-1201 b & 2.49 & 2.42 & 703 & 0.029 & 3476 & 0.51 & 95 & \textbf{10} & Confirmed \\
TOI-1231 b & 24.25 & 3.65 & 330 & 0.129 & 3553 & 0.48 & \textbf{97} & 2 & Confirmed \\
TOI-1242.01 & 0.38 & 2.06 & 1825 & 0.009 & 4255 & 0.66 & \textbf{64} & \textbf{12} & LP \\
TOI-1254.01 & 1.02 & 8.27 & 1968 & 0.019 & 5451 & 0.95 & \textbf{204} & \textbf{75} & LFP \\
TOI-1264.01 & 2.74 & 7.84 & 1260 & 0.036 & 5040 & 0.83 & \textbf{184} & \textbf{55} & pFP \\
TOI-1266 c & 18.80 & 1.67 & 348 & 0.106 & 3573 & 0.45 & \textbf{61} & 0 & Confirmed \\
TOI-1293.01 & 1.68 & 3.71 & 1785 & 0.028 & 5923 & 1.08 & \textbf{63} & 10 & pFP \\
TOI-1347.01 & 0.85 & 2.70 & 1793 & 0.017 & 5300 & 0.96 & \textbf{79} & \textbf{12} & VP \\
TOI-1355.01 & 2.17 & 15.62 & 2900 & 0.043 & 9218 & 2.32 & \textbf{239} & 148 & Inconclusive \\
TOI-1410.01 & 1.22 & 2.94 & 1396 & 0.020 & 4635 & 0.72 & \textbf{118} & \textbf{20} & VP \\
TOI-1444 b & 0.47 & 1.40 & 2324 & 0.012 & 5430 & 0.93 & 5 & \textbf{5} & Confirmed \\
TOI-1468.01 & 15.53 & 2.64 & 337 & 0.086 & 3496 & 0.34 & \textbf{66} & 1 & Confirmed \\
TOI-1468.02 & 1.28 & 1.45 & 682 & 0.021 & 3496 & 0.34 & 10 & \textbf{6} & Confirmed \\
TOI-1518 b & 1.90 & 21.02 & 2492 & 0.039 & 7300 & 1.79 & 197 & \textbf{269} & Confirmed \\
TOI-1546.01 & 1.13 & 3.03 & 2357 & 0.023 & 6223 & 1.21 & \textbf{62} & \textbf{9} & pFP \\
TOI-1634 b & 0.99 & 1.75 & 924 & 0.015 & 3550 & 0.50 & \textbf{79} & \textbf{13} & Confirmed \\
TOI-1683.01 & 3.06 & 2.64 & 929 & 0.037 & 4402 & 0.69 & \textbf{101} & 12 & VP \\
TOI-1715.01 & 2.83 & 5.61 & 1962 & 0.046 & 7072 & 1.58 & \textbf{124} & 22 & LFP \\
TOI-1770.01 & 1.09 & 4.47 & 2447 & 0.022 & 6273 & 1.21 & \textbf{119} & 24 & LFP \\
TOI-1798.02 & 0.44 & 1.41 & 2122 & 0.011 & 5165 & 0.86 & 8 & \textbf{6} & VP \\
TOI-1806.01 & 15.15 & 3.41 & 337 & 0.088 & 3272 & 0.39 & \textbf{60} & 1 & VP \\
TOI-1807 b & 0.55 & 1.26 & 2098 & 0.008 & 4757 & 0.75 & 16 & \textbf{14} & Confirmed \\
TOI-1856.01 & 197.03 & 12.07 & 332 & 0.663 & 5616 & 1.00 & \textbf{59} & 1 & pFP \\
TOI-1895.01 & 748.07 & 9.57 & 276 & 1.921 & 6762 & 1.69 & \textbf{30} & 0 & LP \\
TOI-1954.01 & 4.90 & 7.87 & 1815 & 0.071 & 7120 & 1.97 & \textbf{104} & 23 & Inconclusive \\
TOI-1967.01 & 0.43 & 5.09 & 2981 & 0.011 & 5343 & 1.03 & \textbf{123} & \textbf{39} & LFP \\
TOI-2076 c & 21.02 & 3.50 & 735 & 0.109 & 5187 & 0.82 & 169 & \textbf{13} & Confirmed \\
TOI-2134.01 & 9.23 & 3.10 & 676 & 0.076 & 4406 & 0.69 & \textbf{196} & 16 & VP \\
TOI-2260 b & 0.35 & 1.62 & 2627 & 0.010 & 5534 & 0.99 & 74 & \textbf{9} & Confirmed \\
TOI-2299.01 & 165.02 & 3.69 & 325 & 0.513 & 5780 & 0.66 & \textbf{90} & 1 & pFP \\
TOI-2324.01 & 1.04 & 2.49 & 2560 & 0.022 & 6413 & 1.30 & \textbf{61} & \textbf{8} & LP \\
TOI-2341.01 & 0.88 & 8.59 & 1075 & 0.015 & 3495 & 0.61 & 208 & \textbf{104} & pFP \\
TOI-2407.01 & 2.70 & 3.79 & 718 & 0.031 & 3596 & 0.52 & 78 & \textbf{11} & LP \\
TOI-2411 b & 0.78 & 1.68 & 1358 & 0.014 & 4099 & 0.65 & 69 & \textbf{10} & Confirmed \\
TOI-2427 b & 1.31 & 1.80 & 1114 & 0.020 & 4072 & 0.64 & 131 & \textbf{17} & Confirmed \\
TOI-2455.01 & 4.72 & 14.33 & 576 & 0.043 & 3553 & 0.49 & \textbf{280} & 83 & LFP \\
TOI-2590.01 & 0.75 & 1.78 & 2425 & 0.017 & 6162 & 1.17 & \textbf{57} & \textbf{6} & LFP \\
TOI-2640.01 & 0.91 & 7.39 & 705 & 0.012 & 2999 & 0.25 & 179 & \textbf{56} & LFP \\
TOI-2673.01 & 1.91 & 2.37 & 1739 & 0.030 & 5601 & 0.97 & 70 & \textbf{9} & LP \\
TOI-3235.01 & 2.59 & 11.40 & 604 & 0.027 & 3389 & 0.39 & 162 & \textbf{106} & Confirmed \\
TOI-3353.01 & 4.67 & 2.67 & 1264 & 0.060 & 6365 & 1.33 & \textbf{90} & 8 & VP \\
TOI-3884.01 & 4.54 & 6.00 & 462 & 0.035 & 3269 & 0.28 & \textbf{460} & 24 & Confirmed \\
TOI-4317.01 & 238.85 & 3.78 & 240 & 0.662 & 4354 & 0.68 & \textbf{25} & 0 & LP \\
TOI-4336.01 & 16.34 & 2.11 & 318 & 0.085 & 3365 & 0.31 & \textbf{89} & 1 & LP \\
TOI-4337.01 & 2.29 & 2.82 & 880 & 0.029 & 3953 & 0.62 & \textbf{123} & \textbf{15} & LFP \\
TOI-4340.01 & 2.67 & 3.46 & 1832 & 0.041 & 6406 & 1.28 & \textbf{49} & 6 & LP \\
TOI-4443.01 & 1.85 & 1.72 & 1639 & 0.030 & 5834 & 1.05 & \textbf{97} & 8 & VP \\
TOI-4481.01 & 0.93 & 1.33 & 942 & 0.014 & 3600 & 0.41 & 44 & \textbf{24} & Confirmed \\
TOI-4495.01 & 5.18 & 3.63 & 1383 & 0.062 & 6156 & 1.17 & \textbf{74} & 9 & VP \\
TOI-4506.01 & 5.41 & 2.89 & 332 & 0.034 & 2938 & 0.17 & \textbf{176} & \textbf{3} & LFP \\
TOI-4524.01 & 0.93 & 1.72 & 2140 & 0.018 & 5596 & 1.01 & \textbf{46} & 10 & Confirmed \\
TOI-4527.01 & 0.40 & 0.91 & 1363 & 0.008 & 3702 & 0.48 & 32 & \textbf{13} & VP \\
TOI-4537.01 & 6.66 & 3.87 & 1115 & 0.071 & 5975 & 1.10 & \textbf{150} & \textbf{16} & LFP \\
TOI-4552.01 & 0.30 & 1.28 & 1128 & 0.006 & 3304 & 0.27 & \textbf{22} & \textbf{20} & Inconclusive \\
TOI-4597.01 & 4.67 & 13.23 & 1801 & 0.067 & 7712 & 1.83 & \textbf{532} & 145 & Inconclusive \\
TOI-4602.01 & 3.98 & 2.55 & 1380 & 0.051 & 6012 & 1.12 & \textbf{111} & \textbf{11} & VP \\
TOI-4641.01 & 22.10 & 9.77 & 1040 & 0.174 & 6798 & 1.44 & \textbf{237} & 41 & LP \\
TOI-4644.01 & 0.32 & 1.78 & 1989 & 0.008 & 4242 & 0.66 & \textbf{65} & 11 & LP \\
TOI-4666.01 & 2.91 & 12.87 & 744 & 0.033 & 3666 & 0.58 & \textbf{192} & 76 & LP \\
TOI-4670.01 & 15.50 & 10.83 & 334 & 0.086 & 3382 & 0.35 & \textbf{97} & \textbf{4} & pFP \\
TOI-4856.01 & 14.49 & 7.56 & 339 & 0.080 & 3376 & 0.33 & \textbf{84} & 3 & LFP \\
TOI-4860.01 & 1.52 & 8.58 & 695 & 0.018 & 3255 & 0.34 & 180 & \textbf{88} & Confirmed \\
TOI-5019.01 & 1.09 & 8.97 & 1970 & 0.020 & 5479 & 0.97 & 94 & \textbf{35} & LFP \\
TOI-5023.01 & 2.27 & 17.05 & 581 & 0.023 & 3199 & 0.30 & \textbf{230} & \textbf{152} & LFP \\
TOI-5082.01 & 4.24 & 2.55 & 1165 & 0.051 & 5670 & 1.00 & \textbf{160} & \textbf{15} & VP \\
TOI-5118.01 & 1.57 & 2.51 & 2331 & 0.030 & 6635 & 1.39 & \textbf{33} & \textbf{4} & pFP \\
TOI-5135.01 & 2.02 & 3.47 & 2093 & 0.032 & 5810 & 1.04 & \textbf{66} & \textbf{11} & LFP \\
TOI-5179.01 & 0.29 & 5.53 & 2997 & 0.009 & 5702 & 1.02 & \textbf{154} & \textbf{46} & pFP \\
TOI-5268.01 & 2.07 & 8.27 & 585 & 0.021 & 3162 & 0.29 & \textbf{209} & \textbf{46} & LFP \\
TOI-5278.01 & 0.44 & 21.09 & 1111 & 0.008 & 3349 & 0.36 & \textbf{568} & \textbf{702} & pFP \\
TOI-5311.01 & 39.05 & 13.53 & 345 & 0.196 & 3793 & 0.66 & \textbf{56} & 3 & LFP \\
TOI-5315.01 & 2.47 & 17.17 & 643 & 0.027 & 3333 & 0.43 & 162 & \textbf{111} & pFP \\
TOI-5319.01 & 4.08 & 3.75 & 602 & 0.039 & 3580 & 0.49 & \textbf{111} & \textbf{11} & pFP \\
TOI-5367.01 & 1.66 & 9.79 & 1792 & 0.029 & 6155 & 1.18 & 88 & \textbf{28} & LFP \\
TOI-5388.01 & 2.59 & 1.89 & 601 & 0.024 & 3495 & 0.29 & \textbf{185} & \textbf{12} & VP \\
TOI-5394.01 & 15.19 & 9.68 & 847 & 0.124 & 5977 & 1.10 & \textbf{373} & 62 & pFP \\
TOI-5425.01 & 0.46 & 7.91 & 2860 & 0.013 & 6439 & 1.29 & 114 & \textbf{39} & LFP \\
TOI-5486.01 & 2.02 & 3.79 & 818 & 0.026 & 3654 & 0.54 & 87 & \textbf{15} & LP \\
TOI-5575.01 & 32.0736228 & 9.26 & 219 & 0.197 & 3176 & 0.21 & \textbf{193} & 0 & pFP \\
TOI-5579.01 & 19.86 & 7.82 & 346 & 0.108 & 3595 & 0.43 & \textbf{80} & 3 & LFP \\
TOI-5695.01 & 2.22 & 9.36 & 596 & 0.023 & 3157 & 0.35 & \textbf{217} & \textbf{55} & LFP \\
TOI-5735.01 & 2.12 & 0.86 & 528 & 0.018 & 3222 & 0.18 & \textbf{19} & 2 & pFP \\
TOI-5746.01 & 711.76 & 12.63 & 162 & 1.411 & 4662 & 0.74 & \textbf{78} & 0 & LFP \\
TOI-5747.01 & 0.57 & 1.50 & 1080 & 0.010 & 3466 & 0.41 & 19 & \textbf{17} & LP \\
TOI-5792.01 & 0.70 & 7.01 & 2327 & 0.016 & 5726 & 1.03 & \textbf{208} & \textbf{68} & LFP \\
TOI-5800.01 & 2.63 & 2.83 & 1133 & 0.034 & 4821 & 0.79 & \textbf{164} & \textbf{21} & pFP \\
TOI-5806.01 & 3.19 & 10.72 & 1801 & 0.047 & 6602 & 1.38 & \textbf{387} & 118 & pFP \\
TRAPPIST-1 b & 1.51 & 1.12 & 397 & 0.012 & 2566 & 0.09 & \textbf{28} & 4 & Confirmed \\
TRAPPIST-1 c & 2.42 & 1.10 & 340 & 0.016 & 2566 & 0.09 & \textbf{24} & 2 & Confirmed \\
TRAPPIST-1 d & 4.05 & 0.79 & 286 & 0.022 & 2566 & 0.09 & \textbf{26} & 0 & Confirmed \\
TRAPPIST-1 e & 6.10 & 0.92 & 250 & 0.029 & 2566 & 0.09 & \textbf{20} & 0 & Confirmed \\
TRAPPIST-1 f & 9.21 & 1.05 & 218 & 0.038 & 2566 & 0.09 & \textbf{17} & 0 & Confirmed \\
TRAPPIST-1 h & 18.77 & 0.76 & 172 & 0.062 & 2566 & 0.09 & \textbf{16} & 0 & Confirmed \\
WASP-18 b & 0.94 & 13.35 & 2438 & 0.020 & 6432 & 1.29 & 25 & \textbf{319} & Confirmed \\
WASP-19 b & 0.79 & 15.64 & 2047 & 0.017 & 5440 & 0.96 & 156 & \textbf{224} & Confirmed \\
WASP-29 b & 3.92 & 6.79 & 973 & 0.046 & 4800 & 0.77 & 97 & \textbf{70} & Confirmed \\
WASP-33 b & 1.22 & 18.82 & 2735 & 0.024 & 7308 & 1.50 & \textbf{463} & \textbf{579} & Confirmed \\
WASP-34 b & 4.32 & 14.04 & 1160 & 0.052 & 5700 & 0.96 & \textbf{348} & \textbf{171} & Confirmed \\
WASP-39 b & 4.06 & 14.93 & 1150 & 0.049 & 5485 & 0.93 & \textbf{450} & 98 & Confirmed \\
WASP-43 b & 0.81 & 11.28 & 1465 & 0.014 & 4500 & 0.58 & 103 & \textbf{355} & Confirmed \\
WASP-69 b & 3.87 & 11.85 & 959 & 0.045 & 4700 & 0.98 & \textbf{743} & \textbf{215} & Confirmed \\
WASP-76 b & 1.81 & 20.51 & 2182 & 0.033 & 6250 & 1.46 & \textbf{484} & \textbf{334} & Confirmed \\
WASP-77 A b & 1.36 & 13.56 & 1720 & 0.023 & 5617 & 0.90 & 188 & \textbf{351} & Confirmed \\
WASP-80 b & 3.07 & 11.05 & 824 & 0.034 & 4143 & 0.58 & 295 & \textbf{217} & Confirmed \\
WASP-94 A b & 3.95 & 19.28 & 1498 & 0.055 & 6153 & 1.67 & \textbf{578} & 170 & Confirmed \\
WASP-107 b & 5.72 & 10.54 & 733 & 0.055 & 4425 & 0.68 & \textbf{959} & 93 & Confirmed \\
WASP-121 b & 1.27 & 19.65 & 2454 & 0.026 & 6776 & 1.36 & \textbf{317} & \textbf{268} & Confirmed \\
WASP-127 b & 4.18 & 14.70 & 1422 & 0.048 & 5620 & 0.95 & \textbf{851} & 138 & Confirmed \\
WASP-144 b & 2.28 & 9.53 & 1269 & 0.032 & 5200 & 0.81 & 69 & \textbf{55} & Confirmed \\
WASP-166 b & 5.44 & 7.06 & 1273 & 0.064 & 6050 & 1.19 & \textbf{231} & 36 & Confirmed \\
WASP-189 b & 2.72 & 18.15 & 2638 & 0.051 & 8000 & 2.03 & \textbf{301} & 378 & Confirmed
\enddata

\end{deluxetable}

%% file: tfop_obs_table_short.tex
Table \ref{tab:tfop_obs} contains information on all of the publicly-available \ac{TFOP} observations that were used in the synthesis of the reconnaissance photometry and spectroscopy dispositions incorporated into our vetting and validation analysis. We note that no photometry or spectroscopy data were used directly, only the synthesized dispositions created for each target by \ac{TFOP}'s SG1 and SG2. Also contained in Table \ref{tab:tfop_obs} is information on the high-resolution imaging that was used as observational constraints in our statistical validation analysis. A further overview of how these dispositions and follow-up observations were used can be found in Sections \ref{sec:vetting} and \ref{sec:validation}.

It is important to note that not included in this table are observations that are not publicly-available on ExoFOP\footnote{\url{https://exofop.ipac.caltech.edu/tess/}}. This may be becayse the observations area currently within a proprietary period, contained within a private archive (e.g. an archive accessible only to members of a specific collaboration), or the observing team decided not to post their observations publicly. However, there was a subset of observations that fall under this category that were communicated to the TFOP sub-group leads for use in the synthesis of dispositions and were therefore indirectly utilized by our analysis, but cannot be listed here as they were not made public on ExoFOP. These include observations from HARPS-N (TOI-261.01, TOI-1683.01, TOI-5082.01), PFS (TOI-261.01), Minerva-Asutralis (TOI-261.01), CHIRON (TOI-1895.01), NRES (TOI-3353.01, TOI-5082.01), CORALIE (TOI-3353.01), and Keck/HIRES (TOI-1683.01). We direct the reader to these teams for further information on the observations obtained by these instruments that are not publicly-available on ExoFOP.

Furthermore, not every observation that is currently posted on ExoFOP was utilized in the synthesis of TFOP dispositions for each target. This is because an observer needs to submit their observations to the respective sub-group in an opt-in fashion for the observations to be used in the synthesis of TFOP dispositions.

\startlongtable
\begin{deluxetable}{ccccc}
\tabletypesize{\footnotesize}
\tablecaption{Follow-up observations used in synthesis of TFOP dispositions that were incorporated into our vetting and validation analysis. A full, machine-readable version of this table is available in the online version of this article.}
\label{tab:tfop_obs}
\tablehead{ \colhead{TOI} & \colhead{Telescope} & \colhead{Camera/Instrument} & \colhead{Filter/Bandpass} & \colhead{Observation Date} }

\startdata
\multicolumn{5}{c}{\textit{SG1 Photometry}} \\
128.01 & LCO-CTIO & SBIG 0.4m & ip & 11/14/18 \\
 & Solaris SLR2 & Andor Ikon-L & V & 9/20/18 \\
 & LCO-SSO & Sinistro & zs & 10/15/18 \\
 & LCO-SSO & SBIG 0.4m & ip & 1/12/19 \\
212.01 & El Sauce & SBIG STT1603-3 & unfiltered & 11/16/18 \\
 & Hazelwood Observatory & STT3200 & Rc & 11/17/18 \\
 & PEST & ST-8XME & Ic & 11/23/18 \\
 & InfraRed Survey Facility & SIRIUS & J,H,Ks & 11/15/18 \\
 & KELT South & Apogee Instruments AP16E & Kodak Wratten No. 8 (Rk) & 4/26/10 \\
 & TRAPPIST-South & FLI ProLine PL3041-BB & z & 11/27/18 \\
 & El Sauce & SBIG STT1603-3 & Ic & 12/1/18 \\
\multicolumn{5}{c}{\vdots} \\
\multicolumn{5}{c}{\textit{SG2 Spectroscopy}} \\
128.01 & SMARTS & CHIRON & 4500-8900 \AA & 2/15/19 \\
 & SMARTS & CHIRON & 4500-8900 \AA & 4/7/19 \\
 & SMARTS & CHIRON & 4500-8900 \AA & 1/7/21 \\
179.01 & SMARTS & CHIRON & 4500-8900 \AA & 2/18/19 \\
& SMARTS & CHIRON & 4500-8900 \AA & 8/13/19 \\
& SMARTS & CHIRON & 4500-8900 \AA & 2/2/20 \\
& SMARTS & CHIRON & 4500-8900 \AA & 2/5/20 \\
& SMARTS & CHIRON & 4500-8900 \AA & 2/10/20 \\
& SMARTS & CHIRON & 4500-8900 \AA & 12/21/20 \\
238.01 & FLWO & TRES & 3850-9096 \AA & 12/20/18 \\
 & SMARTS & CHIRON & 4500-8900 \AA & 5/23/19 \\
\multicolumn{5}{c}{\vdots} \\
\multicolumn{5}{c}{\textit{SG3 High-resolution Imaging}} \\
128.01 & VLT & NaCo & Brgamma & 12/16/18 \\
 & Gemini & DSSI & R & 10/31/18 \\
179.01 & Gemini & Zorro & 562 nm & 9/12/19 \\
 & Gemini & Zorro & 832 nm & 9/12/19 \\
 & Gemini & Zorro & 562 nm & 1/12/20 \\
 & Gemini & Zorro & 832 nm & 1/12/20 \\
212.01 & VLT & NaCo & Ks & 1/25/19 \\
 & Gemini & Zorro & 562 nm & 10/8/22 \\
 & Gemini & Zorro & 832 nm & 10/8/22 \\
 & Gemini & Zorro & 562 nm & 7/29/22 \\
 & Gemini & Zorro & 832 nm & 7/29/22 \\
 \multicolumn{5}{c}{\vdots} \\
\enddata

\end{deluxetable}